\documentclass{ptephy_v1}

\preprintnumber{XXXX-XXXX} 
\usepackage{hyperref}
 \usepackage{ulem}

\newcommand{\blm}[1]{{\mbox{\boldmath $#1$}}}
\newcommand{\wada}[1]{\textcolor{black}{#1}}
\newcommand{\eng}[1]{\textcolor{black}{#1}}

\begin{document}

\title{Radiative Acceleration and X-ray Spectrum of Outflowing Pure Electron-Positron Pair Fireball in Magnetar Bursts}


\author[1,2,3]{Tomoki Wada}
\affil[1]{Department of Physics, National Chung Hsing University, Taichung, 40227, Taiwan
\email{tomoki.wada@astr.tohoku.ac.jp}}
\affil[2]{Frontier Research Institute for Interdisciplinary Sciences and Astronomical Institute, Graduate School of Science, Tohoku University, Sendai, Miyagi, 980-8578, Japan} 
\author[3]{Katsuaki Asano}
\affil[3]{Institute for Cosmic Ray Research, The University of Tokyo, Kashiwa, Chiba, 277-8582, Japan}


\begin{abstract}
An X-ray short burst associated with a Galactic fast radio burst 
was observed in 2020,
distinguished by its X-ray cut-off energy significantly exceeding that of other X-ray 
short bursts. X-ray photons of these short bursts are believed to originate from fireballs 
within the magnetospheres of magnetars. If a fireball forms near a magnetic pole, 
it expands along the magnetic field lines, subsequently emitting photons and generating 
plasma outflows that may account for the observed radio bursts. 
We numerically study the radiative acceleration and X-ray spectrum of such outflowing fireballs
consisting of pure electron-positron pairs and radiation, employing spherically symmetric relativistic
radiation hydrodynamics calculations 
with the effects of strong magnetic fields.
Using Monte-Carlo scheme in the radiation calculation, we 
consistently incorporate both the acceleration of the fluid by radiation and the scattering 
of radiation by the fluid, both of which are enhanced by the cyclotron resonant scattering.
Our calculation reveals that cyclotron resonant scattering accelerates the plasma outflow
significantly and broadens the X-ray spectrum. The plasma outflow is accelerated up to ultra-relativistic 
velocities, with Lorentz factors exceeding 100. The calculated X-ray spectrum broadened due to the 
scattering is similar to the observed X-ray spectrum in the Galactic fast radio burst.
\end{abstract}
\subjectindex{E15,E32,E35,E38}

\maketitle

\section{Introduction} \label{sec:int}
Magnetars are neutron stars with X-ray luminosities exceeding their spin-down luminosities 
\citep[e.g.,][for reviews]{HarLai2006,KasBel2017,EnoKis2019_neutronstar}.
Their spin-down luminosities indicate that the magnetars possess strong magnetic
field\eng{s} of $\sim 10^{14}$--$10^{15}\,{\rm G}$, which are 
\eng{thought to be the primary energy source}
of the 
X-ray luminosities \citep{DunTho1992}.
Magnetars 
exhibit bursting activities
\eng{such as}
short bursts, outbursts, and flares. 
The duration of short bursts and flares is roughly 
1 s, and they are categorized 
\eng{based on their isotropic}
luminosity 
(\eng{$\sim10^{36}$--$10^{43}\,{\rm erg\,s^{-1}}$} for the short bursts, and 
\eng{$\sim 10^{44}$--$10^{47}\,{\rm erg\,s^{-1}}$}
for the flares; \cite{KasBel2017}).
The short bursts and the flares are considered to be triggered by a sudden
release of their magnetic energy \citep{ThoDun1995,ThoDun1996}.
The released energy is promptly thermalized; an electron-positron 
pair plasma tightly coupled with photons, 
\eng{whose total energy exceeds that of the pair plasma,} 
so-called a fireball, is formed.
X-ray photons of short bursts and flares are emitted from this fireball.

\eng{The dynamics of fireballs} and the X-ray spectrum depend on the shape of the magnetic 
field at the largest scale \eng{around the fireball} 
and on the ratio of the thermal energy density to 
the magnetic energy density.
(1) Trapped fireball: If 
\eng{a} fireball is formed far from the magnetic pole and the thermal energy 
density is much lower than the magnetic energy density, the fireball 
\eng{is likely to be confined} in a closed magnetic flux tube.
\eng{If this fireball is confined in a closed magnetic field lines of 
a dipole magnetic field, the}
photon energy 
\eng{emitted} from the trapped fireball is typically $\sim 10\,{\rm keV}$ 
\cite[see][for details]{ThoDun1995}. The X-ray spectrum of the emission
from the trapped fireball has been studied, and some observed X-ray spectra 
of short bursts 
\eng{have been successfully}
reproduced \citep{Lyu1986,Lyu2002,YamLyu2020}.
The polarization of 
X-ray emission is also studied in \cite{YanZha2015}.
(2) Expanding fireball: If the thermal energy density 
\eng{of a fireball}
is much higher than the magnetic energy density, the fireball 
\eng{is not} confined in the magnetic flux tube and 
expand\eng{s almost} unaffected by the magnetic field.
This type of fireball 
\eng{is thought to be} realized in the initial spike of the magnetar flares 
\citep{ThoDun1995,ThoDun2001}. In this case, the X-ray spectrum is 
\eng{approximately} blackbody, which agrees with the observed magnetar flare
properties \citep{HurCli1999,HurBog2005,BogZog2007}.
(3) Outflowing fireball:
\eng{If} the released energy is much lower than the magnetic energy, an energy 
release near the magnetic pole may form an outflowing fireball
\eng{\cite[see][for the detailed scenarios]{ThoDun2001,Iok2020,YanZha2021}}. The fireball 
confined in the open magnetic flux tube would expand along it \citep{ThoDun2001}.
For this case, the X-ray spectrum and polarization are numerically studied in 
\cite{PutWat2016} under a given non-relativistic fluid profile.

The expanding/outflowing fireball in magnetar short bursts may be the source of 
fast radio bursts (FRBs). FRBs are bright radio bursts with a duration of a few 
milliseconds \citep{LorBai2007,Tho2013}. The number of detected FRBs so far exceeds
800, but their source and the emission mechanism of coherent radio waves are still 
unclear \citep[][for reviews]{Kat2018,CorCha2019_FRB,PlaWel2019,PetHes2019,Zha2020_rev,Zha2020_mechanism,Lyu2021_emission,PetHes2021_FRB}.
The origins of almost all FRBs are extragalactic, but there are 
\eng{exceptions, including} FRB 20200428A
\citep{Boc2020,CHIME2020_200428}. FRB 20200428A originated from a Galactic magnetar,
SGR1935+2154, and was associated with 
\eng{a short burst, whose X-ray spectrum is harder than that of other bursts from this magnetar} \citep{Mer2020,Li2021,Rid2021,Tav2021,LiGe2022}.
This event implies that at least a fraction of FRBs accompany X-ray short bursts, but the 
physical connection between FRBs and X-ray bursts, even the energy budget, is still under
debate \citep[e.g.,][]{LuKum2020,Kat2020,Iok2020,YanZha2021,YamKas2022}. 

FRBs may be emitted from a relativistically outflowing plasma driven by expanding/outflowing 
fireballs \citep{YanZha2021,WadIok2023}. In an optically thick region, the relativistic outflow 
is accelerated by an isotropic radiation pressure. This acceleration mechanism has been studied 
in the fireball model of gamma-ray bursts
[\citealp{CavRee1978,Goo1986,Pac1986,ShePir1990} and \citealp{Pir1999,Zha2018G} for reviews].
Even in an optically thin region, if the radiation luminosity is much higher than the kinetic 
luminosity of the plasma \eng{outflow, this}
plasma outflow is further accelerated by radiative force 
\citep{MesLag1993,GriWas1998,NakPir2005,ChoLaz2018}.
In addition to the fireball model of the gamma-ray bursts, this acceleration mechanism has 
been discussed in the context of the relativistic outflow from active galactic nuclei as well
\citep{IwaTak2004,AsaTak2007,AsaTak2009}.
\eng{In magnetar busts,}
X-ray photons, responsible for the radiative acceleration,
can be resonantly scattered at the radius where the 
photon energy equals the cyclotron
energy determined by the magnetic field. Thus, this resonant scattering 
should be taken into account to evaluate the dynamics of the relativistic plasma outflow 
\citep{WadIok2023}.

In this paper, we numerically demonstrate that plasma outflows are accelerated up to an 
ultra-relativistic speed (a Lorentz factor of ${\mathcal O}(10^2)$) by the radiative force
via cyclotron resonant scattering, and 
\eng{derive} photon spectra similar to the observed X-ray spectrum of the short burst 
associated with the FRB 20200428A.
We solve the radiation hydrodynamic equations numerically under the assumption that the
fireball is composed of an electron-positron pair plasma and the flow is spherically symmetric 
and steady. 
\wada{
Wada and Ioka (2023, hereafter WI23)\cite{WadIok2023} 
shows that a ``pure" electron-positron pair fireball, aided by 
radiative acceleration via resonant scattering, can provide sufficient kinetic energy 
to account for the observed FRBs, within a factor of 2 uncertainty. Furthermore, the 
baryon loading on the fireball makes it even easier to account for the energy budget of FRBs.}
The Lorentz factor of the baryon-loading fireball is lower than the pair fireball due to the 
higher mass density, but only by a factor of \eng{approximately} 2 \wada{as long as the photon 
luminosity is an order of magnitude higher than the kinematic luminosity of the plasma 
(Figure 5 in WI23).}
WI23 also derives analytical expressions for the terminal Lorentz factor 
of the outflows. In this paper, we verify their analytical solutions with numerical simulations.
We solve both the radiation and \eng{the} plasma outflows and provide steady solutions for them.

This paper is organized as follows.
In Section~\ref{sec:met}, we describe the method and the boundary condition for the numerical simulations.
In Section~\ref{sec:res}, we present results of the numerical simulations.
In Section~\ref{sec:radacc}, we show that the calculated X-ray spectrum is within the error
range of the observed X-ray spectrum under a 
\eng{suitable} boundary condition, and a relativistic
plasma outflow is also created by the resonant scattering.
In Section~\ref{sec:par}, we show the parameter dependences of 
\eng{the photon spectrum and the radiative acceleration, which is analytically estimated in Section~\ref{sec:ana}.}
Section~\ref{sec:con} is devoted to the conclusions and discussion.
We use $Q_x= Q/10^x$ in cgs units unless otherwise noted.

\section{Method}\label{sec:met}
\subsection{Physical Setup}\label{sec:phy}
\begin{figure}
  \center
  \includegraphics[width=0.5\linewidth]{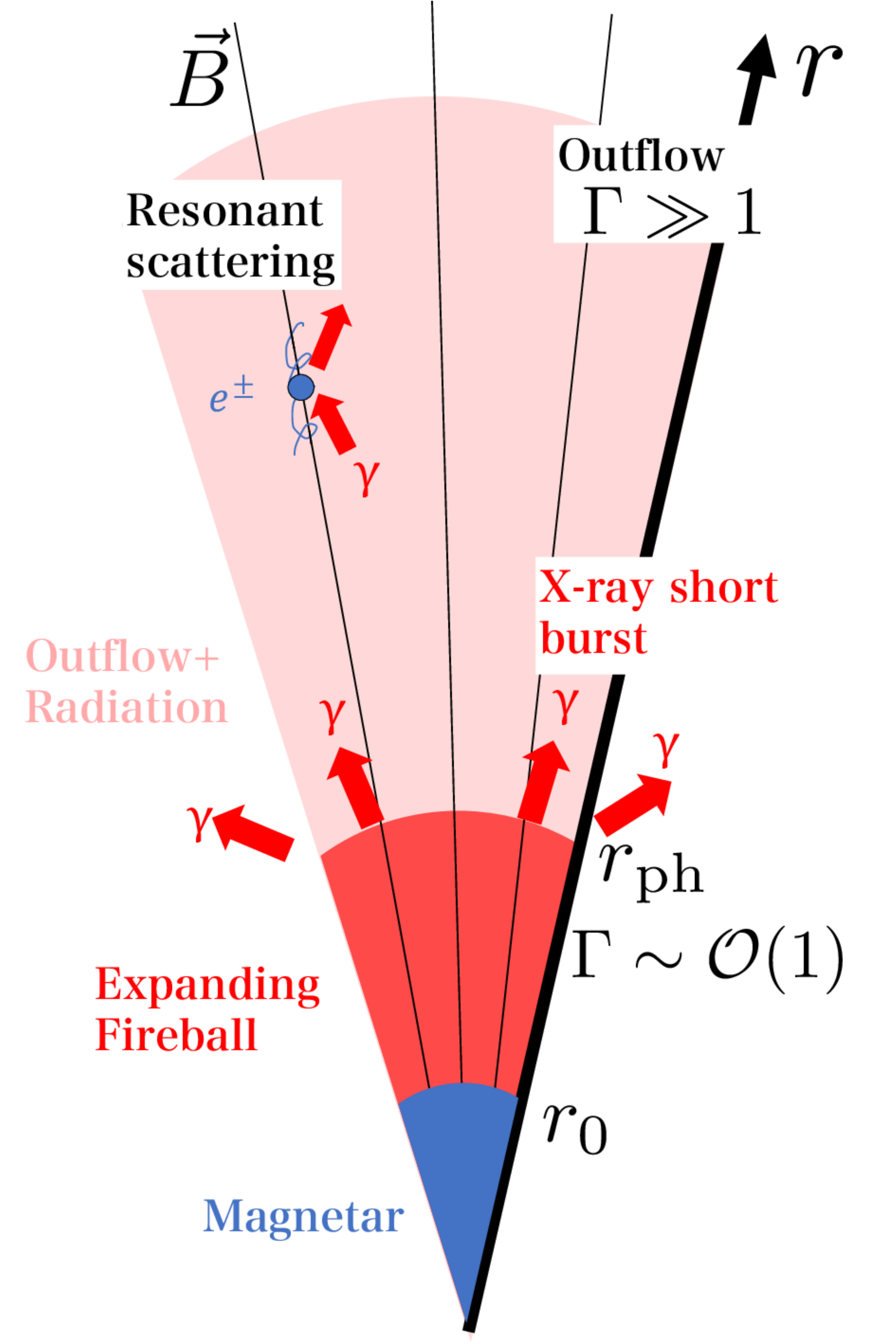}
  \caption{
    Schematic picture of the outflowing pair fireball.
    From the magnetar surface at $r_0$, an optically thick fireball expands \eng{outward} 
    to the Thomson photospheric radius, $r_{\rm ph}$.
    \eng{Above this radius,} pair creation and annihilation are almost negligible,
    and almost all photons are not scattered via the Thomson scattering.
    We 
    \eng{perform} 
    numerical simulations from this radius.
    As photons propagate \eng{outward}, the magnetic field decreases, and the photons 
    undergo 
    resonant scattering where
    \eng{their} frequency 
    \eng{matches} the cyclotron frequency.
    Photons are Comptonized 
    \eng{through} the resonant scattering, 
    \eng{and}
    as the backreaction of this scattering, the outflow is accelerated.
    \label{fig:abst}
  }
\end{figure}

An outflowing fireball along an open flux tube is 
\eng{generated} in a magnetar short burst, 
\eng{if substantial} energy is 
\eng{deposited} at the 
\eng{base} of the open magnetic field lines.
\eng{This energy deposition} may 
\eng{originate} from a trapped fireball
\eng{confined within} the closed magnetic field lines,
\eng{preventing its expansion}\citep{ThoDun2001, Iok2020,YanZha2021,DemLyu2023}.
This trapped fireball is formed by crustal shear oscillations or magnetic reconnections,
\eng{which release} 
thermal energy in a 
\eng{localized} region \citep[e.g.,][]{ThoDun1995,Lyu2003}.
The total energy of the outflowing fireball and the initial size,
\eng{defined by its initial solid angle,} are determined by the 
the magnetic field \eng{geometry} 
\eng{at} the energy injection \eng{cite.}
In this study, 
\eng{these parameters are treated as inputs.}
In \eng{a} pure electron-positron pair plasma at the temperature
\eng{considered herein,}
the energy density of the fireball is dominated by radiation 
\eng{rather than} electron-positron pairs.
As the fireball expands along the flux tube, it 
\eng{undergoes adiabatic cooling, causing the plasma number density in the thermal equilibrium decrease.
Ultimately,}
the fireball becomes optically thin
\eng{with respect to} Thomson scattering.
\wada{Photons initially escape} around 
\eng{the} Thomson photospheric radius, $r_{\rm ph}$, where the optical depth for 
Thomson scattering equals unity (see Figure~\ref{fig:abst}).
\eng{It is noted} that $r_{\rm ph}$ is defined for the 
standard Thomson cross section in this 
\eng{study.}
The escaped photons 
will be 
observed as X-ray short bursts.

Above the Thomson photospheric radius, the pair plasma and radiation are no longer 
in thermal equilibrium, and the radiation streaming almost freely is \eng{anisotropic.}
The plasma is accelerated by \eng{the momentum transfer from the radiation via scattering.}
If the \eng{photon} luminosity
\eng{significantly exceeds} the kinetic luminosity of the plasma, 
the number density of photons is much larger than that of plasma. 
\eng{While} most photons 
\eng{remains unscattered by the plasma, almost all}
plasma particles are continuously scattered by photons.
\eng{Consequently,} the plasma is subjected to the radiative force and is accelerated \citep{MesLag1993,GriWas1998,LiSar2008,AsaTak2009}.
\eng{This force is proportional to the product of the photon flux and the cross section
\citep[e.g., Sec 6.4 of][]{MihalasMihalas1984}.}

In \eng{the presence of} a magnetic field, photons can be resonantly scattered 
with 
cross section \eng{exceeding}
the Thomson cross section.
This cyclotron resonant scattering occurs at 
\eng{radii} where \eng{the condition} $\omega\sim\omega_B$ is satisfied,
\eng{where}
$\omega$ is the angular frequency of a photon, $\omega_B= eB/mc$ is the cyclotron frequency,
$e$ is the elementary charge, $B$ is the magnetic field, $m$ is the electron mass, and $c$ 
is the speed of light.
Because of the
\eng{enhanced} cross section, the radiative force via 
resonant scattering is stronger than 
\eng{one via} Thomson scattering
\eng{allowing} the plasma outflow 
\eng{to achieve} higher Lorentz factors.
\eng{Furthermore,} the photon spectrum can be 
\eng{altered} by the resonant scattering above $r_{\rm ph}$ \citep[e.g.,][]{YamLyu2020}.

We
\eng{investigate} the radiative acceleration 
\eng{beyond} the Thomson photospheric radius and the \eng{resulting} Comptonized photon
spectrum consistently assuming a spherically symmetric steady flow.
Below the Thomson photospheric radius, photons and electron-positron pairs are 
\eng{tightly} coupled, and follow the hydrodynamic equations with an adiabatic index of $4/3$.
Around the Thomson photospheric radius, the pair annihilation freezes out, and the numbers of
photons and electron-positron pairs 
\eng{remain almost constant} above this radius \citep[e.g.][]{AsaTak2009}.
\eng{Beyond} the Thomson photospheric radius, photons follow the radiative transfer equation 
and the electron-positron pairs follow the hydrodynamic equations. 
\eng{By solving these equations} iteratively,
we obtain steady-state solutions.

In this paper, we consider fireballs composed of electron-positron pairs, 
\eng{excluding}
baryons 
for simplicity.
\eng{The inclusion of baryons in the fireball introduces two primary differences.}
First, the terminal \wada{four-}velocity of the outflow 
\eng{decreases compared to} the result of this paper.
Second, the eigenmodes of electromagnetic waves in the plasma with baryons are circularly 
polarized, while the eigenmodes in the pair plasma are linearly polarized. \eng{Moreover,}
the scattering cross sections
\eng{corresponding to} these modes 
\eng{differ, potentially altering} 
the propagation of photons.
The baryon-loaded fireball will be studied in our future papers.

Photons follow the radiation transfer equation with the scatterings in a magnetized electron-positron 
pair plasma, \eng{expressed as follows:}
\cite[e.g.,][]{Lin1966,MihalasMihalas1984}\cite[][for mathematical formulation]{Sas1958,Sas1962}
\begin{eqnarray}
 p^\mu\left(\frac{\partial}{\partial x^\mu}-\Gamma^\rho_{\mu\nu}p^\nu\frac{\partial}{\partial p^\rho}\right)F_\gamma(x^\mu,{\mbox{\boldmath $p$}})=\left(\frac{dF_\gamma}{d\lambda}\right)_{\rm coll},\label{eq:boltzmann}
\end{eqnarray}
where $\blm{p}$ is the spacial component of the four-momentum of photons $p^\mu$, $x^\mu$ 
\eng{represents} the spacetime coordinates, $\Gamma^\rho_{\mu\nu}$ is the Clistoffel symbol,
$F_\gamma(x^\mu,\blm{p})$ is the distribution function of photons, 
$\left(\frac{dF_\gamma}{d\lambda}\right)_{\rm coll}$ is the collision term, and $\lambda$ is 
the affine parameter.
We solve Equation~(\ref{eq:boltzmann}) 
\eng{using a} Monte-Carlo scheme in the plasma flow obtained by the fluid simulation.
\eng{The equations are formulated in covariant notation for ease of adaptation to polar coordinates.}

The energy-momentum conservation is \eng{as follows:}
\begin{eqnarray}
\nabla_\mu T^{\mu\nu}+\nabla_\mu T_{\rm rad}^{\mu\nu}=0,
  \label{eq:eom}
\end{eqnarray}
where $\nabla_\mu$ is covariant derivative with respect to $x^\mu$, 
$T^{\mu\nu}$ is the energy-momentum tensor of anisotropic fluid, and 
$T_{\rm rad}^{\mu\nu}$ is that of radiation. For given radiation field,
$-\nabla_\mu T_{\rm rad}^{\mu\nu}$ represents the radiative force on the
plasma, \eng{evaluated using the numerical solution to}
Equation~(\ref{eq:boltzmann}).
Neglecting electron-positron pair annihilation/creation, the mass flux 
conservation is written as:
\begin{eqnarray}
  \nabla_\mu (\rho u^\mu)=0,
  \label{eq:eoc}
\end{eqnarray}
where $\rho$ is the comoving 
\wada{rest-mass density}
of plasma, and 
$u^\mu$ is the four-velocity of plasma (see Section~\ref{sec:hyd} for details).
We solve Equations~(\ref{eq:eom}) and (\ref{eq:eoc}) under the radiative
force obtained by the radiation simulation.

For simplicity, we assume a spherically symmetric steady flow and neglect
the gravity of the magnetar. \eng{The magnetic field is presumed to be radial,}
\eng{and photon absorption and emission are assumed to be negligible} 
above the Thomson photospheric radius. \eng{Below this radius, the}
the fireball 
is \eng{treated as } a thermalized electron-positon pair plasma and photons
(see Section~\ref{sec:ini}).

\subsection{Hydrodynamics of Anisotropic Plasma}\label{sec:hyd}
\wada{
Due to the significant synchrotron cooling, electrons and positrons predominantly occupy 
the ground state of the Landau levels;
consequently, the equations of motion 
governing a relativistic pair outflow 
must incorporate the anisotropic pressure.}
The energy-momentum tensor of anisotropic plasma is 
\eng{expressed} as follows \citep{CheGol1956,Ged1991,GedOib1995},
\begin{eqnarray}
T^{\mu\nu}=\wada{\rho c^2  h} u^\mu  u^\nu +P_\perp g^{\mu\nu}+\left(P_\parallel-P_\perp\right)\hat{b}^\mu\hat{b^\nu},
\end{eqnarray}
where $h$ is the specific enthalpy of the fluid, $P_\perp$ is the pressure perpendicular to magnetic field,
$P_\parallel$ is the pressure parallel to magnetic field, $g^{\mu\nu}$ is the metric of spacetime, and 
$\hat{b}^\mu$ 
\eng{represents} a tetrad \eng{aligned with the }
magnetic field in the comoving frame of the fluid.
The fluid flows along the radial flux tube, and the magnetic pressure does not exert any force on the fluid.
The specific enthalpy is expressed as
\begin{equation}
 h=1+\frac{e_{\rm th}}{\rho \wada{c^2}}+\frac{P_\perp}{\rho\wada{c^2}},
\end{equation}
where $e_{\rm th}$ is the thermal energy of the fluid.
In the case that the synchrotron cooling timescale is 
\eng{significantly} shorter than the timescale for \eng{the plasma} isotropization,
particles are in the ground state of the Landau levels, 
\eng{leading to}
\begin{equation}
    P_\perp =0.\label{eq:Pperp}
\end{equation}
\eng{Thus}, the energy-momentum tensor 
\eng{for} the anisotropic plasma in this case is \eng{given by}
\begin{eqnarray}
 T^{\mu\nu}&=&\rho \wada{c^2}h u^\mu  u^\nu +P_\parallel\hat{b}^\mu\hat{b}^\nu.\label{eq:Tmunu}
\end{eqnarray}
\wada{Furthermore, in the parameter 
regime examined in this 
study, the speed of thermal motion of pairs parallel to the magnetic field lines is 
substantially slower than the speed of light.}
We solve Equation~(\ref{eq:eom}) 
\eng{employing} this energy-momentum tensor \wada{and an equation of state for non-relativistic anisotropic gas}.

For the spherically symmetric flow, we 
\eng{utilize} polar coordinates $(\wada{ct},r,\theta,\phi)$, where the components of metric \eng{are given by}
\begin{eqnarray}
  g_{\mu\nu}={\rm diag}(\wada{-1},1,r^2,r^2\sin^2 \theta),
\end{eqnarray}
and then, the components of $u^\mu$ and $\hat{b}^\mu$ are \eng{represented as}
\begin{eqnarray}
u^\mu&=&(\Gamma,\Gamma\beta,0,0),\label{eq:u}\\    
\hat{b}^\mu&=&(\Gamma\beta,\Gamma,0,0),\label{eq:hatb}
\end{eqnarray}
where $\beta$ is the radial velocity of the fluid normalized by the speed of light, 
and $\Gamma=(1-\beta^2)^{-1/2}$.
\wada{The four-velocity $u^\mu$ is normalized to satisfy $u^\mu u_\mu=-1$.} 
\eng{By} substituting Equations~(\ref{eq:u}) and 
(\ref{eq:hatb}) into Equation~(\ref{eq:Tmunu}), the components of the energy-momentum 
tensor 
\eng{for} the plasma are \eng{given by}
\begin{eqnarray}
T^{tt}&=&\rho \wada{c^2}h_\parallel \Gamma^2-P_\parallel,\label{eq:Ttt}\\
T^{tr}&=&\rho \wada{c^2}h_\parallel \Gamma^2\beta,\label{eq:Ttr}\\
T^{rr}&=&\rho \wada{c^2}h_\parallel \Gamma^2\beta^2+P_\parallel,\\
T^{ij}&=&0~({\rm others}),\label{eq:Tij}
\end{eqnarray}
\eng{in which} we define
\begin{eqnarray}
  h_\parallel=1+\frac{e_{\rm th}}{\rho\wada{c^2}}+\frac{P_\parallel}{\rho\wada{c^2}},
  \label{eq:hpara}
\end{eqnarray}
which is the analogy of the specific enthalpy of an isotropic fluid.
We employ the $\Gamma$-law ideal equation of state,
\begin{eqnarray}
  P_{\parallel}=(\hat{\Gamma}-1)e_{\rm th},
  \label{eq:eos}
\end{eqnarray}
where $\hat{\Gamma}$ is the adiabatic index.
For the gas 
\eng{wherein} the thermal motion of all particles is \wada{non-relativistic} 
and in one dimension \eng{(aligned with the magnetic field line in this context),}
$\hat{\Gamma}$ 
\eng{takes on the value of} $3$.
Using Equations~(\ref{eq:eom}), (\ref{eq:eoc}), (\ref{eq:u}), and (\ref{eq:Ttt})--(\ref{eq:Tij}), 
the continuity equation and the equation of motions 
\eng{can be expressed as}
\begin{eqnarray}
  \partial_t (r^2\rho \Gamma)+\partial_r(r^2\rho\Gamma\beta)&=&0,\label{eq:eocc}  \\
  \partial_{t}\left[r^2\left(\rho \wada{c^2} h_\parallel \Gamma^2-P_\parallel\right)\right]
  +\partial_{r}\left[r^2\rho \wada{c^2}h_\parallel \Gamma^2\beta\right]
  &=&r^2 G_{\rm rad}^t,\label{eq:eomt}\\
  \partial_{t}\left[r^2\rho \wada{c^2}h_\parallel \Gamma^2\beta\right]
  +\partial_{r}\left[r^2\left(\rho \wada{c^2}h_\parallel \Gamma^2\beta^2
  \wada{+}P_\parallel\right)\right]
  &=&r^2 G_{\rm rad}^r,\label{eq:eomr}
\end{eqnarray}
where 
\wada{$\partial_t$ and $\partial_r$} denote the partial 
\eng{derivative} with respect to 
\wada{$ct$ and $r$}, and $G_{\rm rad}^\mu=-\nabla_\nu T_{\rm rad}^{\mu\nu}$
\eng{represents} the radiation four-force density, which is 
calculated 
\eng{based on} the result of the Monte-Carlo calculation (see Section~\ref{sec:rad}).
\wada{In contrast to the case of an fluid with an isotropic pressure, 
the terms originating from $T^{\theta\theta}$ and $T^{\phi\phi}$ are
not present in Equation~(\ref{eq:eomr}), 
since $T^{\theta\theta}$ and $T^{\phi\phi}$ are proportional to $P_\perp$
\wada{for radial outflows} and $P_\perp=0$ here.}
We solve Equations~(\ref{eq:eocc}) -- (\ref{eq:eomr}) numerically for $\rho$, $\Gamma$, and $P_\parallel$.
The parallel temperature \wada{in the fluid comoving frame} $T_\parallel$ satisfies \citep{Ged1991},
\begin{eqnarray}
    \wada{P_\parallel=\frac{\rho}{m}T_\parallel\label{eq:prt},}
\end{eqnarray}
\wada{where the temperature is measured in the dimension of energy.}
The perpendicular temperature \wada{in the fluid comoving frame}, $T_\perp$, equals zero due to
the strong synchrotron cooling (see Equation~\ref{eq:Pperp}).

The numerical method and setup employed in our study are \eng{outlined} as follows.
We adopt the relativistic HLLE approximate Riemann solver \cite{Ein1988,SchKat1993}\cite[][for a review]{MarJos2003}
and 2nd-order MUSCL reconstruction for the spatial interpolation \citep{Van1979} with the minmod function as a flux
limiter \citep{Roe1986}. The time integral is performed with the second-order Runge-Kutta method. Our code \eng{has} successfully
passe\eng{ed} 
\eng{several} shock tube tests in \cite{SchKat1993}. We set the CFL number 
\eng{in the range of} $10^{-4}$--$10^{-3}$ because the 
radiative force is larger than that of the energy-momentum flux of the fluid (see Equations~\ref{eq:eomt} and 
\ref{eq:eomr}). The radius is divided into about $9000$ cells. We check that our simulation result 
\eng{remain consistent at}
a higher-resolution simulation, in which the cell size is half. To conduct this simulation with a small CFL number, we 
divide the space into 3--5
\eng{segments. Calculations for each segment are performed sequentially from the innermost to outermost regions until 
steady} 
solutions are realized in each 
\eng{section}. As we aim to obtain steady solutions, it does not matter if the transient behavior 
i.e., the time-dependence of the solution is somewhat imprecise.

\subsection{Radiation Transfer in a Magnetized Plasma}\label{sec:rad}
The 
photon scattering is 
\eng{incorporated} as the collision term in Equation~(\ref{eq:boltzmann}).
\eng{In the absence of}
absorption or emission process, the collision term is 
\eng{expressed} as \citep{Lin1966}
\begin{eqnarray}
  \left(\frac{dF_\gamma}{d\wada{\lambda}}\right)_{\rm coll}&=&\frac{\rho(x)}{\wada{m}}\left[-\kappa(x,p)F_\gamma(x^\mu,\blm{p})
  \wada{+}\int d\Pi^\prime \kappa(x,p^\prime)\zeta(x;p^\prime\to p)F_\gamma(x^\mu,\blm{p}^\prime)\right]
  \label{eq:scat}
\end{eqnarray}
where $\rho(x)/\wada{m}$ is the comoving number density of plasma, $\kappa(x,p)\rho(x)/\wada{m}$ is a invariant 
scattering opacity, $\zeta(x;p^\prime\to p)$ is a invariant phase function \citep[e.g.,][]{Chandrasekhar1960}, 
and $d\Pi=p^t\,dp^t\,d\Omega$ for photons in a flat spacetime. We denote the solid angle element in momentum space with $d\Omega$.
The first term \wada{in the square brackets on} the right-hand side of Equation~(\ref{eq:scat}) 
\eng{represents} the 
decrease of the number of photons due to the scattering out of the momentum $\blm{p}$, 
and the second term 
\eng{describes} the increase due to the scattering from a momentum $\blm{p}^\prime$ to 
$\blm{p}$.
The function $\kappa(x,p)$ and the scattering cross section satisfy
\begin{eqnarray}
 \kappa(x,p)= p^{t}\sigma\left(x,p\right),
 \label{eq:kappadef}
\end{eqnarray}
where $\sigma\left(x,p\right)$ is the cross section in the frame where $p^t$ is measured.
The invariant phase function satisfies
\begin{eqnarray}
 \zeta(x;p^\prime\to p)=\frac{1}{p^t\sigma\left(x,p^{\prime}\right)}\frac{d\sigma}{dp^td\Omega}.
\end{eqnarray}
The specific expression of the differential cross-section 
\eng{depends on} the plasma physics in a magnetic field.

Electromagnetic waves in an electron-positron pair plasma can be decomposed 
into two eigenmodes of linearly polarized waves: the extraordinary mode (X-mode) and the 
ordinary mode (O-mode) \citep[e.g.,][]{Lifshitz1981,Meszaros1992}. The electric field of 
an O-mode photon 
\eng{lies in} the plane 
\eng{containing} the wavenumber vector of the photon and the magnetic field,
\eng{whereas} the electric field of an X-mode photon is perpendicular to that plane. 
In the frame where the particle motion
parallel to the magnetic field 
\eng{is} zero (hereafter denoted as the electron rest frame\footnote{
\eng{Strictly} speaking, the electron is not at rest due to the zero-point 
oscillation in the ground state of the Landau level but we neglect it.}),
the cross sections 
\eng{for} an X-mode photon and an O-mode photon
are written as \citep[e.g.,][]{CanLod1971,Her1979,Ven1979,Nag1981,Meszaros1992,YanZha2015}
\begin{eqnarray}
  \frac{d\sigma^{\rm X X} }{d\Omega' }
  &=&\frac{3\sigma_{\rm T}}{32\gamma_e}\left[ \frac{\bar{\gamma}_e/\pi}{\left( 1+\mathfrak{u}^{1/2}\right) ^{2}+\bar{\gamma}_e ^{2}}+\frac{\bar{\gamma}_e/\pi}{\left( 1-\mathfrak{u}^{1/2}\right)^{2}+\bar{\gamma}_e^{2}}\right], \label{eq:sXX}\\
  \frac{d\sigma^{\rm X O} }{d\Omega' }
  &=&\frac{3\sigma_{\rm T}}{32\gamma_e} \cos^2\theta^\prime\left[ \frac{\bar{\gamma}_e/\pi}{\left( 1+\mathfrak{u}^{1/2}\right) ^{2}+\bar{\gamma}_e ^{2}}+\frac{\bar{\gamma}_e/\pi}{\left( 1-\mathfrak{u}^{1/2}\right)^{2}+\bar{\gamma}_e^{2}}\right],\label{eq:sXO}\\
  \frac{d\sigma^{\rm O X} }{d\Omega' }
  &=&\frac{3\sigma_{\rm T}}{32\gamma_e}\cos^2\theta\left[ \frac{\bar{\gamma}_e/\pi}{\left( 1+\mathfrak{u}^{1/2}\right) ^{2}+\bar{\gamma}_e ^{2}}+\frac{\bar{\gamma}_e/\pi}{\left( 1-\mathfrak{u}^{1/2}\right)^{2}+\bar{\gamma}_e^{2}}\right],\label{eq:sOX}\\
  \frac{d\sigma^{\rm OO} }{d\Omega' }
  &=&\frac{3\sigma_{\rm T}}{8\pi}\left\{\sin ^{2}\theta \sin ^{2}\theta '+\frac{\pi}{4\gamma_e} \cos ^{2}\theta \cos ^{2}\theta '\right.\nonumber\\
    &~&\times \left.\left[ \frac{\bar{\gamma}_e/\pi}{\left( 1+\mathfrak{u}^{1/2}\right) ^{2}+\bar{\gamma}_e ^{2}}+\frac{\bar{\gamma}_e/\pi}{\left( 1-\mathfrak{u}^{1/2}\right)^{2}+\bar{\gamma}_e^{2}}\right]\right\},  \label{eq:sOO}
\end{eqnarray}
where $d\sigma^{\rm AB}/d\Omega'~(A,B=X,O)$ is the differential scattering 
cross section of an A-mode photon into a B-mode, $\sigma_{\rm T}$ is the 
Thomson cross section, $\theta$ is the angle between the wavevector of the
incident photon and the magnetic field in the electron rest frame, $\theta'$
is the angle between the wavevector of the scattered photon and the magnetic
field in the electron rest frame, 
$\mathfrak{u} \equiv \mathfrak \omega_B^2/\omega^2$, and $\gamma_e=2e^2\omega_B/(3 m c^3)$
is the radiative damping factor. In the exact formulae, 
$\bar{\gamma}_e=\gamma_e$\eng{; however}
$\bar{\gamma}_e$ \eng{is treated} as a parameter to 
\eng{artificially broaden} 
the resonance profile
(see Sec \ref{sec:rtau}). Here, we neglect the thermal motion of electrons for 
simplicity 
\eng{under the assumption of the cold plasma approximation, where} the temperature
of electrons is lower than the rest-mass energy. The cross sections of 
Equations~(\ref{eq:sXX})--(\ref{eq:sOO}) are used for the collision term of 
Equation~(\ref{eq:scat}).

In the anisotropic plasma
\eng{described} above, the momentum distribution of electron-positron 
pairs is \eng{confined to} one dimensional as \wada{\cite[see][for Monte-Carlo sampling of particles]{CanHow1987},}
\begin{eqnarray}
  f_{\pm,\rm 1D}(p_\parallel,T_\parallel)\,dp_\parallel
  =\frac{ \exp\left(-\frac{\sqrt{m^2c^4+p_\parallel^2c^2}}{T_\parallel}\right)}{\wada{2}mcK_1(mc^2/T_\parallel)} \,dp_\parallel,\label{eq:1ddis}
\end{eqnarray}
where $p_\parallel$ is the momentum of electrons or positrons parallel to the magnetic field,
$T_\parallel$ is the kinetic temperature of pairs parallel to the magnetic field, and $K_1(x)$
is the modified Bessel function of the second kind. 
\wada{The function $f_{\pm,\rm 1D}(p_\parallel,T_\parallel)$ is normalized to satisfy 
$\int_{-\infty}^\infty dp_\parallel\,f_{\pm,\rm 1D}(p_\parallel,T_\parallel)=1$.
The temperature $T_\parallel$ is evaluated 
using $P_\parallel$ and $\rho$, obtained from the fluid simulation, by using Equation~(\ref{eq:prt}).}

In our Monte-Carlo scheme, the radiation field is described by a set of packets 
\citep[e.g.,][]{PozSob1983,DolGam2009,Laz2016,KawFuj2023}. Each packet 
\eng{contains} information 
\eng{regarding} the position ($r$ in one-dimensional simulation), the momentum
($p^t,p^r,p^\theta$), and the polarization \eng{mode} (X-mode or O-mode).
\eng{The computational domain} is divided into spherical cells, and the statistical
quantities of radiation (the intensity, the radiative force, and so on) are recorded 
\eng{in} each cell.

The radiative force $G_{\rm rad}^\nu$ is evaluated from the distribution function, 
$F_\gamma(x^\mu,\blm{p})$ (see \cite{Lin1966} and Appendix~\ref{sec:der}) as
\begin{eqnarray}
  G_{\rm rad}^\nu
  &=&-\frac{\rho(x)}{\wada{m}}h^\nu_\lambda\int d\Pi\, p^\lambda \kappa(x,p) F_\gamma(x^\mu,\blm{p}),\label{eq:radforce}
\end{eqnarray}
where $h^{\nu}_\mu=\delta^\nu_\mu+u^\nu u_\mu$ is the projection tensor onto the 3-space 
orthogonal to $u^\mu$ and $\delta^\nu_\mu$ is the Kronecker delta. In this paper, the temperature of the pair plasma is non-relativistic, and 
the collision term in Equation~(\ref{eq:radforce}) \eng{is evaluated} using 
Equations~(\ref{eq:sXX})--(\ref{eq:sOO}) without taking 
the thermal motion of the plasma into account. In 
Lindquist (1966) \cite{Lin1966}, the expression of 
the radiative force, Equation~(\ref{eq:radforce}), is derived under the assumption 
\eng{of an elastic and isotropic scattering.}
As shown in Appendix~\ref{sec:der}, Equation~(\ref{eq:radforce}) is valid in the 
case that the scattering is elastic and the differential scattering 
cross section is an even function of the direction of the incident and scattered photon.
Because Equations~(\ref{eq:sXX})--(\ref{eq:sOO}) are even functions of $\cos\theta$ and
$\cos\theta^\prime$, Equation~(\ref{eq:radforce}) is applicable. Another method to 
evaluate the radiative force is 
calculate\eng{ing} the radiative force 
\eng{using} the four-momentum changes in all scattering processes in 
the Monte-Carlo simulation\citep[e.g.,][]{RyaDol2015,KawFuj2023}. For steady and strongly 
radiation-dominated cases, 
\eng{the evaluation using} Equation~(\ref{eq:radforce}) 
reduces Monte-Carlo shot noise
because the radiative force 
is evaluated using all photon packets and the statistical error is determined 
by the number of the packets 
not by the number of the scattering. Therefore, we adopt Equation~(\ref{eq:radforce}) 
\eng{in evaluating} the radiative force.

The scattering process is treated as follows. For a packet in a radial cell, 
the probability of escaping the cell is \citep{AbrNov1991}
\begin{eqnarray}
  {\mathcal P}_{\rm esc}&=&1-\exp(-\Delta\tau),\\
  \Delta\tau&=&\Gamma(1-\beta\cos\theta_L)\frac{\rho}{\wada{m}} \sigma^A \Delta l,
  \label{eq:dtau}
\end{eqnarray}
where $\cos\theta_L$ is the angle between the radial direction and the 
wavevector of the packet in the lab frame, $\sigma^A=\sum_{B}\sigma^{AB}$ 
is the cross section for the packet of $A$-mode photons and $\Delta l$ is
the distance that the packet must propagate before escaping from the cell
in the lab frame.\footnote{
The scattering optical depth is calculated as the product of the invariant scattering opacity and the Affine
parameter \citep[e.g.][]{RyaDol2015}, \eng{both of which are}
Lorentz invariant. \eng{In the fluid comoving frame,} the invariant scattering
opacity of a photon is $p^{(t)}\sigma^A\rho/\wada{m}$,
\wada{where $p^{(t)}$ is the energy of the
photon in the fluid comoving frame} (see Equation~\ref{eq:kappadef}; \eng{neglecting} the thermal 
motion of the plasma).
The affine parameter is $\Delta \lambda=c \Delta t/p^t$ in the lab frame. 
\eng{$p^{(t)}$ and $p^{t}$ are related by the Lorentz transformation leading} to $\Delta \tau$ 
in Equation~(\ref{eq:dtau}),
\eng{consistent with the derivation} in 
Abramowicz et al. (1991) \cite{AbrNov1991}.}
We generate a random variable, $\xi$, uniformly distributing in $(0,1]$.
If $\xi\wada{\le}{\mathcal P}_{\rm esc}$, the packet escapes from the current cell.
If $\xi>{\mathcal P}_{\rm esc}$, the packet is scattered in the cell after it propagates
\begin{eqnarray}
  \Delta l^\prime=-\frac{\Delta l}{\Delta \tau}\ln\left[1-\xi\left(1-\exp(-\Delta\tau)\right)\right],
\end{eqnarray}
along its trajectory.
At the position, we probabilistically 
\eng{select} a pair particle from the one-dimensional distribution 
function (see Equation~\ref{eq:1ddis}, \wada{where $T_\parallel$ is evaluated 
from $P_\parallel$ and $\rho$ obtained from the fluid simulation by using 
Equation~\ref{eq:prt}).} 
\eng{Subsequently,} in the rest frame of the selected particle, we probabilistically
determine both the direction and the mode of the 
\eng{packet after the scattering} using Equations~(\ref{eq:sXX})--(\ref{eq:sOO}).
Finally, we transform the angle and the four-momentum of that scattered packet 
\eng{back} to those in the lab frame, and again calculate the probability
of escaping from the cell with a newly generated random variable $\xi'$. 
We continue these processes until the photon escapes from the computational domain.

The adopted parameters and numerical scheme are summarized as follows.
Equation~(\ref{eq:boltzmann}), 
\eng{incorporating} the scattering term, Equations~(\ref{eq:scat}),
(\ref{eq:sXX})--(\ref{eq:sOO}), are solved numerically 
\eng{using} the Monte-Carlo scheme.
The packet propagation is calculated 
\eng{without solving the geodesic equation; for} a photon propagating 
from $r=r_1$ to $r=r_2$, the radial angle in the lab frame evolves from 
$\theta_{L,1}$ to $\theta_{L,2}$
where $r_1/\sin\theta_{L,2}=r_2/\sin\theta_{L,1}$.
\eng{The simulation employs $10^6$ photon packets.}
We checked that our code reproduces the inverse-Comptonized spectrum (without 
fluid motion and with 3D particle distribution) in \cite{PozSob1983}, and 
reproduces the beaming factor in a relativistic outflow (with isotropic scattering 
and 3D particle distribution) studied in \cite{Bel2011}. \wada{We also confirmed that
the energy-momentum tensor of the radiation (isotropic scattering and 3D 
particle distribution) in a given ultra-relativistic fluid profile reproduces 
analytic solutions.}

To reduce the Monte-Carlo shot noise in the radiative force and 
\eng{reduce} computational costs, we artificially increase the 
radiative damping factor, $\bar{\gamma}_e$, in Equations 
(\ref{eq:sXX})--(\ref{eq:sOO}) by a factor of $300$ keeping the frequency-integrated
cross section constant, i.e., widening the Lorentzian function. We also perform an 
additional calculation in the {\tt fiducial} case (see Section~\ref{sec:res} 
and Table~\ref{tab:cases} for details) with the number of photon packets increased 
by a factor of 10 and the broadening width by a factor of only 3. The difference 
in the width broadening factors of 3 and 300 causes a small error in the \wada{four-}velocity less 
than 5\% (see also Section~\ref{sec:rtau}). The error in the X-ray spectra shape 
is approximately 10\% in 1--300 keV range, 
\eng{increasing to 50\% at 600 keV.}

\subsection{Boundary Condition}\label{sec:ini}
The boundary conditions 
\eng{for solving the coupled} 
fluid and radiation equations 
\eng{are specified} as follows (see Figure~\ref{fig:abst}).
At the magnetar surface, $r=r_0$, we set photon luminosity $L_\gamma$, initial
fireball size $r_0\theta_0$, and the magnetic field at the surface $B_0$.
\eng{The value of} $r_0$ \eng{is fixed at}
$10^6\,{\rm cm}$.
The initial temperature of the fireball, $T_0$ is determined 
\eng{by the ralation}
\begin{eqnarray}
  L_\gamma=\pi(r_0\theta_0)^2a_{\rm rad}T_0^4c\Gamma_0^2,
\end{eqnarray}
where $a_{\rm rad}$ is the radiation constant and $\Gamma_0$ is the initial Lorentz factor.
\eng{Under the assumption} that the sonic point 
\eng{lies} near the surface, the initial velocity of the fireball at the magnetar surface 
is set to the sound velocity of radiation-dominated fluid, $\beta_0=1/\sqrt{3}$.

Initially, the fireball is optically thick and treated as a radiation-dominated 
one-component fluid. For 
\eng{the} given boundary condition, we solve the steady flow of the radiation-dominated
fluid from $r=r_0$ to the Thomson photospheric radius, $r_{\rm ph}$, 
\eng{defined by the condition}
\begin{eqnarray}
  \tau_T=\sigma_Tn_\pm (r_{\rm ph},B(r_{\rm ph}))\frac{r_{\rm ph}}{\Gamma(r_{\rm ph})}=1,
  \label{eq:thin}
\end{eqnarray}
where $n_\pm(r,B)$ is the electron-positron pair number density in thermodynamic
equilibrium \citep[e.g.,][]{ThoDun1995}. 
\eng{Within} this optically thick regime, we neglect the contribution of the rest-mass 
\eng{energy} density of the plasma to the specific enthalpy.
Using the temperature
$T_{\rm ph}$ and the velocity 
$\beta_{\rm ph}$ \eng{at the Thomson photospheric radius}, the boundary conditions 
\eng{for} the anisotropic fluid 
\eng{are determined.} The initial density of the plasma is 
\eng{obtained} by solving Equation~(\ref{eq:thin}) with 
$n_\pm=\rho(r_{\rm ph})/\wada{m}$.
The initial velocity of the plasma, $\beta(r_{\rm ph})$, is $\beta_{\rm ph}$.
The kinetic temperature of the fluid, $T_\parallel$, is set to the temperature of the 
radiation-dominated fluid, i.e., $T_\parallel(r_{\rm ph})=T_{\rm ph}$.
Other boundary conditions are 
\eng{derived} using Equations~(\ref{eq:hpara}), (\ref{eq:eos}), and (\ref{eq:prt}).

The boundary conditions for the photon field 
\eng{are defined} at $r_{\rm ph}$. The photon number flux at $r_{\rm ph}$ is \eng{given by}
\begin{eqnarray}
\dot{N}_{\gamma}  =4\pi f_{\rm b} r_{\rm ph}^2\frac{2\bar{\zeta}(3)}{\pi^2} \left(\frac{T_{\rm ph}}{\hbar c}\right)^3 c \Gamma(r_{\rm ph})^2,
\end{eqnarray}
where $\hbar$ is the reduced Planck constant, $N_\gamma$ is the total number of photons,
$\bar{\zeta}(z)$ is the zeta function, and $f_{\rm b}$ is the beaming factor for photons
\eng{expressed} as the ratio of the solid 
\eng{angles subtended by} the polar angle $\theta_0$ 
\eng{and} $\theta_0+\Gamma(r_{\rm ph})^{-1}$:
\begin{eqnarray}
    f_{\rm b}=\frac{\sin^2(\theta_0/2)}{\sin^2[(\theta_0+\Gamma(r_{\rm ph})^{-1})/2]}.
\end{eqnarray}
This beaming factor represents the decrease in the number of photons caused by their 
escaping from the conical region in the simulation due to the weak 
\wada{relativistic beaming effect caused by }
$\Gamma(r_{\rm ph})={\mathcal O}(1)$ at $r=r_{\rm ph}$. 
\wada{By introducing $f_{\rm b}$, the photon loss is taken into 
account at the base of the outflow in our spherically symmetric simulation.}
For the initial photon spectrum, 
we consider two cases. The first 
is the blackbody spectrum with temperature 
$T_{\rm ph}$,
\begin{eqnarray}
  \frac{dn_\gamma}{d\omega}\propto \frac{\omega^2}{\exp(\hbar\omega/T_{\rm ph})-1},
  \label{eq:BB}
\end{eqnarray}
where $n_\gamma$ is the number density of photons. The second 
is the modified blackbody spectrum \citep{Lyu2002},
\begin{eqnarray}
  \frac{dn_\gamma}{d\omega}&\propto&\omega^2 \left[\exp\left(\frac{(\hbar\omega)^2}{T_{\rm ph}\sqrt{(\hbar\omega)^2+(3\pi^2/5)T_{\rm ph}^2}}\right)-1\right]^{-1}.
  \label{eq:modBB}
\end{eqnarray}
The modified blackbody spectrum is realized if the cyclotron energy at $r_{\rm ph}$ 
\eng{significantly exceeds the temperature at the radius}. 
In the case of $\omega\ll\omega_B$, the cross section of the X-mode photons is proportional to
$(\omega/\omega_B)^2$. Thus, photons with lower energy are less frequently scattered and 
escape from the inner region of the fireball with a higher intensity.
As a result, the emergent photon spectrum is softer than the \eng{standard} blackbody spectrum.

In both the two spectra, the initial numbers of photons in X- and O-modes are 
\eng{assumed to be} equal in our simulations, for simplicity. 
\wada{In Lyubarsky (2002) \cite{Lyu2002}, only the X-mode photons are treated to 
derive the modified blackbody spectrum. However, there is great uncertainty in the fireball generation process inside the photosphere, so the actual ratio of O-mode to X-mode photons emitted from a fireball is unknown.  
Since resonant scattering, which we are particularly interested in, occurs in 
both X-mode and O-mode (see Equations~\ref{eq:sXX}--\ref{eq:sOO}), we can
expect no significant modification of the spectra or impact on the dynamics
(see Section~\ref{sec:ana} for analytical estimates of the weak dependence 
of plasma dynamics on the luminosity of photon).}
The initial directions of photons 
\eng{are} randomly 
\eng{sampled} from a uniform distribution 
\eng{over the} solid angle in the comoving frame of the plasma. The initial energies of photons
\eng{are initially distributed uniformly} in logarithmic bins ranging from 
$10^{-15}T_{\rm ph}$ to $10^2T^{\rm ph}$.

The radial dependence of the magnetic field is set as follows. We assume that the 
strength of the magnetic field is proportional to $r^{n_B}$ as
\begin{eqnarray}
  B(r)= B_0\left(\frac{r}{r_0}\right)^{n_B},
  \label{eq:Br}
\end{eqnarray}
where, in terms of the strength (but not direction) of the magnetic field,  
$n_B=-2$ for a split monopole field and $n_B=-3$ for a dipolar field.
Near the magnetic pole, \eng{provided that the angular deviation from the pole 
is much smaller than 1 radian, the $\theta$-component of the dipole magnetic 
field becomes negligible compared to the radial component, resulting in an 
approximately radial outflow.}
We perform the numerical simulation up to $r=1000r_0$, which would be inside the 
light cylinder of the magnetar.
%

The acceleration mechanism discussed below 
\eng{differs fundamentally} from the acceleration of optically thick fireballs
driven by isotropic radiation pressure. \eng{In this case,} the plasma is accelerated by
scattering with photons, most of which freely propagate except around the resonance 
radius. In this case, the strength (not the direction) of the magnetic field along the
flow is the most important factor, which determines the radius at which resonant 
scattering occurs.

\eng{Regarding} the geometrical effect on the radial dependence of the gas pressure,
the contribution of the non-relativistic gas pressure to the energy density in 
Equations~(\ref{eq:eomt}) and (\ref{eq:eomr}) is negligible compared to the
rest-mass energy density.
\eng{Additionally}, the pressure gradient of the gas 
\eng{is insignificant} as it is much 
\eng{weaker} than the radiative force ($r^2G_{\rm rad}^t$, and $r^2G_{\rm rad}^r$ in
Equations~\ref{eq:eomt} and \ref{eq:eomr}). 
\eng{Consequently}, the result of our numerical simulation, 
\eng{particularly} the terminal bulk velocities, 
\eng{are largely unaffected} by the gas pressure evolution.

\section{Results}\label{sec:res}

\begin{table*}[tb]
\begin{center}
 \caption{Parameters in the simulations. 
    $L_\gamma$ is the physical luminosity of the fireball (\textbf{not the isotropic
    luminosity}), $\theta_0$ represents the initial size of the fireball, $B_0$ is 
    the magnetic field at the magnetar surface, $n_B$ represents the radial dependence
    of the magnetic field (see Equation~\ref{eq:Br}).  The ``spectrum'' column indicates
    the initial photon spectrum of the fireball, where ``mod BB'' is the modified 
    blackbody (Equation~\ref{eq:modBB}), and ``BB'' is the blackbody (Equation~\ref{eq:BB}). 
    The ``scattering'' column shows the adopted scattering process in each model;
    ``magnetic'' is the scattering process with the resonant scattering in the magnetic 
    field (see Equations~\ref{eq:sXX}--\ref{eq:sOO}), and ``isotropic'' is the case with 
    only the isotropic Thomson scattering. The radius $r_{\rm ph}$ is the analytic Thomson
    photospheric radius normalized with $r_0$, and $\Gamma_{\rm ph}$ is the Lorentz 
    factor at $r_{\rm ph}$.}
 \label{tab:cases}
  \begin{tabular}{ccccccc|cc}
    model & $L_\gamma\,{\rm [erg\,s^{-1}]}$&$\theta_0$ & $B_0\,{\rm [G]}$ & $n_B$& spectrum&scattering& $r_{\rm ph}$
    & $\Gamma_{\rm ph}$\\
   \hline   
   {\tt fiducial} & $6.0\times10^{39}$        & $3.1\times 10^{-2}$ & $2\times 10^{14}$& -3 & mod BB  &  magnetic   & 1.1 & 1.6\\   
   {\tt iso\_BB} & $6.0\times10^{39}$        & $3.1\times 10^{-2}$ & $2\times 10^{14}$& -3 & BB  &  isotropic  & 1.1 & 1.6\\   
   {\tt BB} & $6.0\times10^{39}$        & $3.1\times 10^{-2}$ & $2\times 10^{14}$& -3 &   BB    &  magnetic & 1.1 & 1.6\\
      {\tt B\_1e16} & $6.0\times10^{39}$        & $3.1\times 10^{-2}$ & $1\times 10^{16}$& -3 &   mod BB      &  magnetic    & 1.2 & 1.8\\   
   {\tt n\_-2}& $6.0\times10^{39}$        & $3.1\times 10^{-2}$ & $2\times 10^{14}$& -2 &     mod BB    & magnetic    & 1.1 & 1.6\\
   \hline
  \end{tabular}
  \end{center}
\end{table*}

We simulate 5 models,
\eng{summarized} in Table~\ref{tab:cases}. In the model {\tt fiducial}, 
the parameters are 
\eng{selected} to reproduce the X-ray spectrum 
\eng{akin} to the observed one, 
\eng{with} the initial spectrum 
\eng{represented by} the modified blackbody 
(see Equation~\ref{eq:modBB}).
\eng{The setups for the other models deviate from {\tt fiducial} as follows:}
In the model {\tt iso\_BB}, we change the initial spectrum into the usual blackbody 
spectrum (see Equation~\ref{eq:BB}), and artificially set the scattering to be an
isotropic scattering with the Thomson cross section 
\eng{disregarding} the effect of the magnetic field.
In the model {\tt BB}, we change only the initial spectrum into the blackbody.
In the model {\tt B\_1e16}, we increase the magnetic field at the surface 
\eng{to} $1\times10^{16}\,{\rm G}$. We chose this extreme magnetic field to 
investigate the effect of varying the magnetic field strength on the 
radiative acceleration. In the model {\tt n\_-2}, we 
\eng{modify} the configuration of the magnetic field from $\propto r^{-3}$ 
to $r^{-2}$ (see Equation~\ref{eq:Br}).

\subsection{Radiative Acceleration and X-ray Spectrum}\label{sec:radacc}

\begin{figure}[tb]
  \begin{center}
                \includegraphics[width=0.7\textwidth]{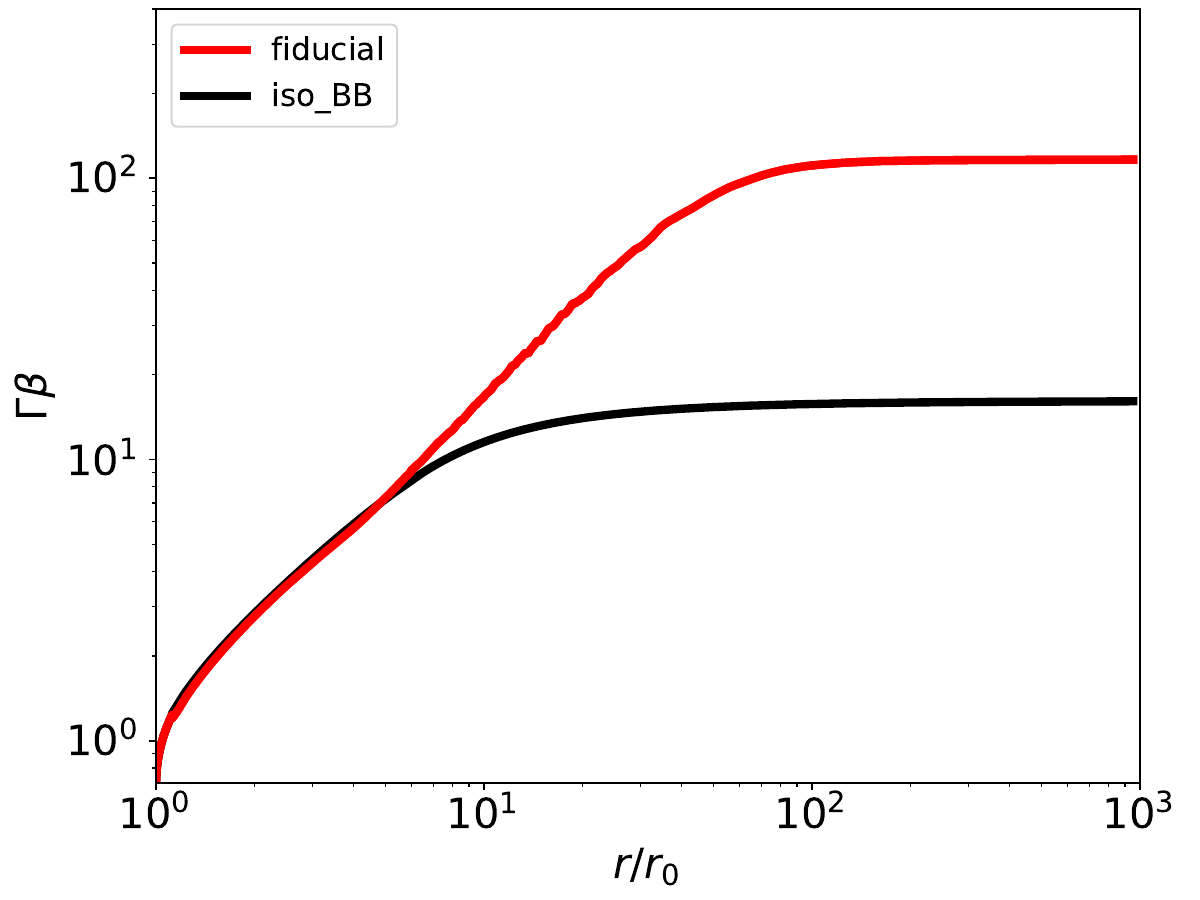}
              \caption{
                \wada{Four-}velocity of the relativistic outflows in the model {\tt fiducial} 
                (red solid line) and the model {\tt iso\_BB} (black solid line).
                The vertical axis is the four-velocity, and the horizontal axis 
                is the radius normalized by the magnetar radius.
                The kinetic luminosity of the plasma shows a similar 
                radial dependence (see Equations~\ref{eq:Ttr} and \ref{eq:eocc} 
                in the cold and steady case, $h_\parallel\simeq1$ and $\partial_t=0$).
                }
              \label{fig:u_temp}
  \end{center}
\end{figure}

Figure~\ref{fig:u_temp} shows the four-velocity \eng{profiles} for the models 
{\tt fiducial} and {\tt iso\_BB}.
The outflow is accelerated by the radiative force even 
\eng{above $r_{\rm ph}$} following \eng{the scaling} $\Gamma\beta \propto r$
\citep{MesLag1993,GriWas1998,LiSar2008}.
In both 
models, the terminal four-velocity 
\eng{significantly exceeds} the \wada{four-}velocity at the Thomson photospheric 
radius, $\Gamma\beta\sim 1$. The acceleration ends at the radius where the work 
\eng{performed} by the radiative force 
\eng{over} the dynamical timescale 
\eng{falls below} the rest mass energy of the outflow (see Section~\ref{sec:ana}
for details). This condition is satisfied around $r/r_0\sim 90$ ($\Gamma\beta\sim 100$) 
in the model {\tt fiducial} (see Equation~\ref{eq:rc_mBB}), and around $r/r_0\sim 10$
($\Gamma\beta\sim 10$) in the model {\tt iso\_BB} (see Equation~\ref{eq:rc_bbiso}).
This 
\eng{discrepancy arises from the exclusion of} the resonant scattering in the model 
{\tt iso\_BB}. The resonant scattering is 
\eng{essential} for 
\eng{producing} the ultra-relativistic outflows with $\Gamma\sim100$ from magnetars.

The outflow at different radii is accelerated by photons with
different \eng{observed} energies, $\hbar\omega_{\rm o}$.
Because the magnetic field is proportional to $r^{n_B}$ (see Equation~\ref{eq:Br}), 
the cyclotron energy is also proportional to $r^{n_B}$. On the other hand, the photon 
energy in the comoving frame of the plasma is $\hbar\omega\sim\hbar\omega_{\rm o}/\Gamma$, 
where $\Gamma$ is proportional to $r$ 
\eng{during the acceleration} and \eng{is} constant 
\eng{during} the coasting phase. 
Photons are resonantly scattered at the radius 
\eng{satisfying} $\hbar eB_0(r/r_0)^{n_B}/(mc)\sim \hbar\omega_{\rm o}/\Gamma$.
During the acceleration, the resonant radius is 
\begin{eqnarray}
\frac{r_{\rm res}}{r_0}\sim \left[\frac{eB_0r_0\Gamma_{\rm ph}}{mc\omega_or_{\rm ph}}\right]^{1/(-n_B-1)}
\sim 5\,B_{0,14.3}^{1/2}(\hbar\omega_{\rm o})_{,100\,{\rm keV}}^{-1/2},\nonumber\\
\label{eq:rres}
\end{eqnarray}
where $(\hbar\omega_{\rm o})_{,100\,{\rm keV}}=\hbar\omega_{\rm o}/100\,{\rm keV}$. 
We have set $\Gamma_{\rm ph}r_0/r_{\rm ph}\sim 1$ for simplicity, and set $n_B=-3$ at 
the last expression. In the similar way, the resonant radius in the coasting phase is 
\begin{eqnarray}
\frac{r_{\rm res}}{r_0}\sim \left[\frac{mc\omega_o}{e B_0\Gamma_t}\right]^{1/n_B}
\sim 100\,B_{0,14.3}^{1/3}(\hbar\omega_{\rm o})_{,0.1\,{\rm keV}}^{-1/3}
\Gamma_{\rm t,2}^{1/3},\nonumber
\label{eq:rresc}\\
\end{eqnarray}
where $\Gamma_{\rm t}$ is the terminal Lorentz factor of the outflow, and 
$\hbar\omega$ is normalized by $0.1\,{\rm keV}$.
In the model {\tt fiducial}, the outflow is accelerated by the radiative force via 
the resonant scattering in $5\lesssim r/r_0\lesssim 50$ (see the difference between
the red line and the black line in Figure~\ref{fig:u_temp}).
\eng{Photons with energies within} 
$100\,{\rm keV}\gtrsim\hbar \omega_o\gtrsim 6\,{\rm keV}$ sequentially contribute 
to this acceleration. This radial dependence of the cross section suggests that 
the terminal \wada{four-}velocity also 
\eng{influenced by} the photon spectrum (see Section~\ref{sec:par}).

\begin{figure}[tb]
  \begin{center}
                    \includegraphics[width=0.7\textwidth]{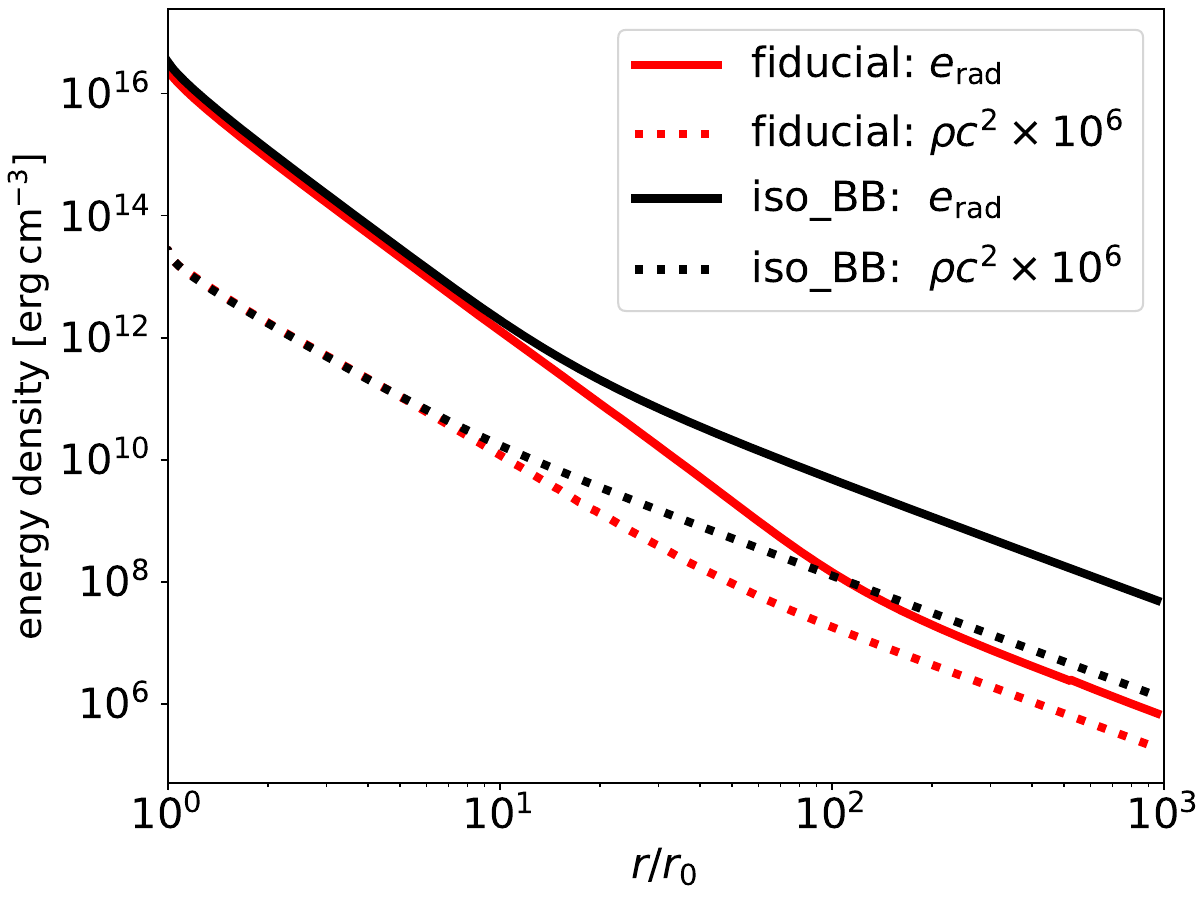}
          \caption{
            Radiation energy density in the comoving frame of plasma (solid lines), 
            $e_{\rm rad}$, and rest mass energy density of plasma in the same frame 
            (dotted lines), $\rho c^2$, in the models {\tt fiducial} (red) 
            and {\tt iso\_BB} (black). The decrease at $r\gtrsim 10r_0$ in the model 
            {\tt fiducial} compared to the model {\tt iso\_BB} is due to the difference 
            in the comoving frame of the plasma (see Figure~\ref{fig:u_temp}).}
          \label{fig:fluid_profile} 
  \end{center}
\end{figure}

Figure~\ref{fig:fluid_profile} shows the radiation energy density and the rest 
mass energy density of plasma in the comoving frame of the plasma.
The radiation energy density in the comoving frame is
$e_{\rm rad}=T^{\mu\nu}_{\rm rad}u_\mu u_\nu$.
The radiation energy density 
\eng{significantly exceeds} the plasma energy density, which is usual for pure 
electron-positron pair fireballs. Note that the density decrease in the 
model {\tt fiducial} for $r\gtrsim 10 r_0$ 
\eng{results from} the acceleration, not the decrease of the photon 
luminosity.\footnote{
Photons are not isotropic in the comoving frame of the plasma for $r\gtrsim 10r_0$.
For example, in the case where photons stream freely in the radial 
direction, $p^t=p^r$ in this frame, and the energy-momentum tensor of the radiation,
$T^{\mu\nu}_{\rm rad}=\int d\Pi\,p^\mu p^\nu F_\gamma(x^\mu,\blm{p})$, satisfies 
$T_{\rm rad}^{tt}=T_{\rm rad}^{tr}=T_{\rm rad}^{rr}$. Then, the radiation energy 
density in the comoving frame is 
$e_{\rm rad}=T_{\rm rad}^{tt}\Gamma^2-2T_{\rm rad}^{tr}\Gamma^2\beta+T_{\rm rad}^{rr}\Gamma^2\beta^2\simeq T^{tt}_{\rm rad}/(4\Gamma^2)$,
which decreases with increasing Lorentz factor.
}
Since the energy density is measured in the 
comoving frame \eng{of the plasma}, $e_{\rm rad}$ in the model {\tt fiducial} 
and that in the model {\tt iso\_BB} are measured in different frames, especially 
in $r\gtrsim10r_0$.

\begin{figure}[tb]
  \begin{center}
              \includegraphics[width=0.7\textwidth]{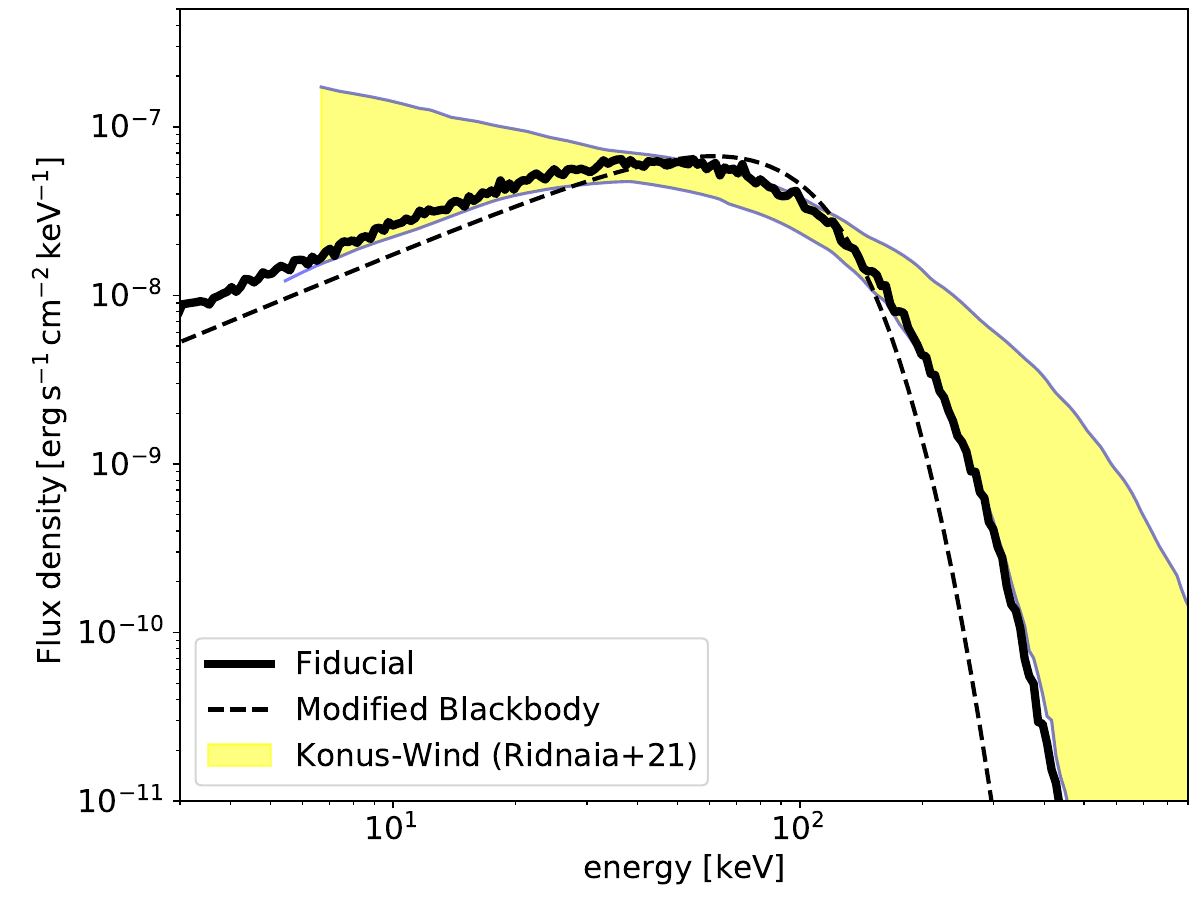}
              \caption{
                X-ray spectrum in 5--900 keV of the model {\tt fiducial} (thick solid).
                The vertical axis is the flux density, and the horizontal axis is the photon energy.
                The black dashed line is the modified blackbody spectrum for reference.
                The luminosity distance to the magnetar is set to $7$ kpc.
                The yellow-shaded region is the 99.7\% error region of the X-ray short burst associated with FRB 20200428A, which is detected from SGR 1935+2154 by the Konus-{\it Wind} \citep{Rid2021}.
              }
              \label{fig:spec}          
  \end{center}
\end{figure}

Figure \ref{fig:spec} shows the X-ray spectrum \wada{calculated from our
spherically symmetric one-dimensional radiation calculation}
in the model {\tt fiducial},
which is in agreement with the observation. 
Above the Thomson photospheric radius, the photons are up-scattered in the 
relativistic outflow, resulting in the X-ray spectrum within the error range 
of the observed one.
The pure modified blackbody spectrum is inconsistent with the observation
(the yellow-shaded region in Figure~\ref{fig:spec})
\wada{at the 3-sigma level.}

\wada{The broadening of the spectrum,
also seen in the simulations in \cite{Pee2008,AsaTak2009,ItoNag2013},
is due to two main reasons. 
One is the superposition of emissions with 
different beaming factors \cite{Goo1986, Pee2008,ItoNag2013}. 
The photons from high latitudes (in other words, off-axis region) have a smaller beaming 
factor than 
the photons from the line of sight (in other words, on-axis), leading to lower 
peak photon energy. These high-latitude components become prominent at energy 
ranges lower than the peak photon energy of the spectrum, $\sim 100\,{\rm keV}$.
}
\wada{Another reason is the scattering in regions where the optical
depth is close to unity. 
The typical $T_\parallel$ at that radius is $\sim 10$ keV. Photons with energies lower than $\sim 10 \Gamma$ keV can gain energy via scattering. Scattered photons contribute to both the high-energy ($\gtrsim 100\,{\rm keV}$) and low-energy ($\lesssim 100\,{\rm keV}$) parts of the spectrum.}

The spectrum may be further broadened by considering the 
\wada{2D or 3D} 
structure of the fireball
\citep{LunPee2013,ItoNag2013,ItoNag2014} or the baryons loading \citep{RydLun2017}, both 
of which are not taken into account in this paper. If these effects are taken 
into account, the X-ray spectrum may be in even better agreement with the observation.

\subsection{Parameter Dependences of the Radiative Acceleration}\label{sec:par}

\begin{figure}[tb]
  \begin{center}
          \includegraphics[width=0.7\textwidth]{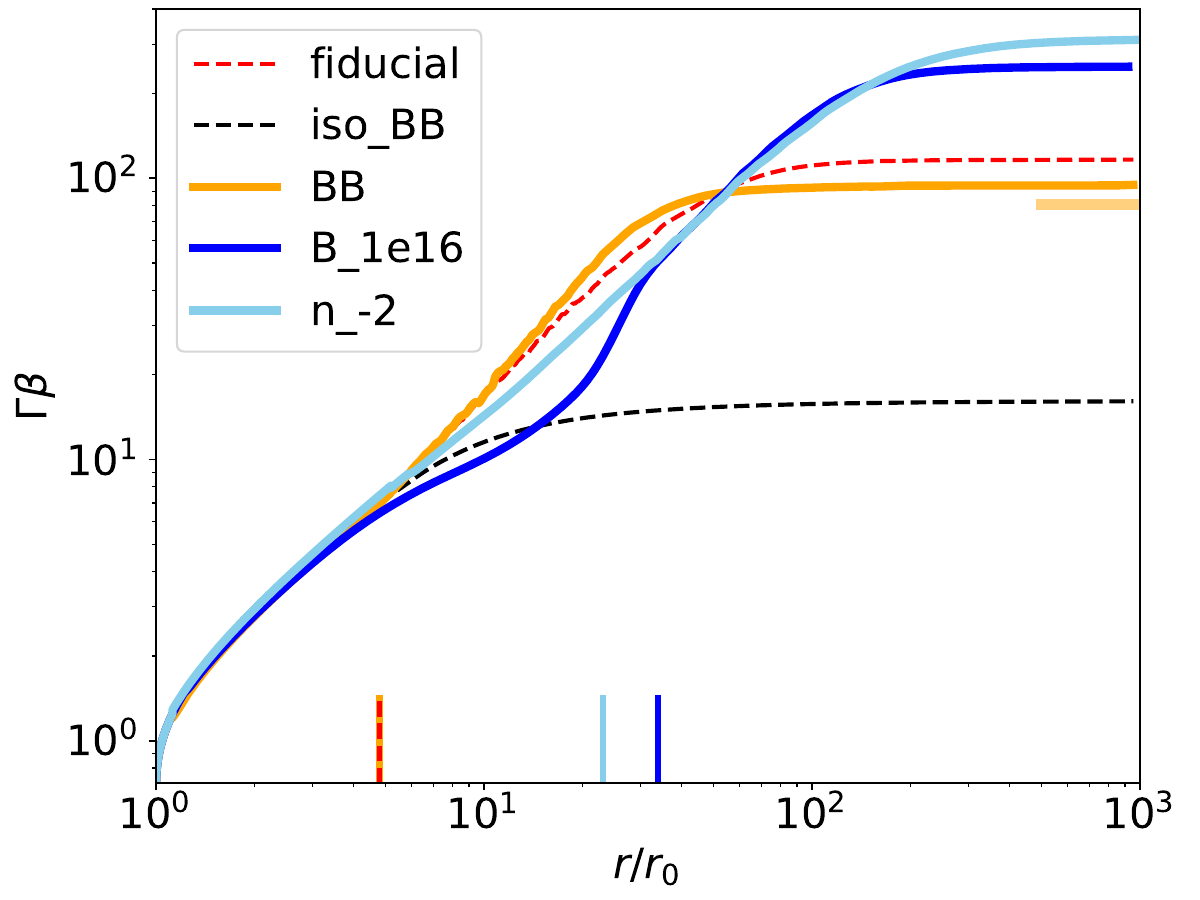}
          \caption{
            \wada{Four-}velocity of the relativistic outflows in all the models. The axes
            are the same as Figure~\ref{fig:u_temp}. The dashed red and black 
            lines are the models {\tt fiducial} and  {\tt iso\_BB}, respectively.
            The solid orange, blue, and sky blue lines are the models {\tt BB},
            {\tt B\_1e16}, and {\tt n\_-2}. The lower vertical line is the 
            resonant radius for the 100 keV photons, where 100 keV photons in 
            the observer frame are last resonantly scattered (see Equation~\ref{eq:rres}).
            The rightmost horizontal orange line is the terminal \wada{four-}velocity derived 
            for the parameters in our simulations using the formula in 
            WI23 (their Equations~52 and 53).
          }
          \label{fig:u_comp}
  \end{center}
\end{figure}

Figure~\ref{fig:u_comp} shows the \wada{four-}velocity of the outflow for all the models 
(see Table~\ref{tab:cases}). The diverse radial dependencies of the \wada{four-}velocity 
\eng{arise from} the resonant scattering and the 
\eng{varietion in} photon spectra. The terminal \wada{four-}velocity of the model {\tt BB} is 
\eng{slightly} lower than that of the model {\tt fiducial}. 
\eng{As the radius increases,}
the cyclotron frequency decreases 
\eng{according to} $\omega_B\propto B\propto r^{-3}$,
\eng{resulting in} lower energy photons 
\eng{being} resonantly scattered and 
\eng{contributin} to the radiative force.
The photon number density in the low-energy region ($\omega\ll T_{\rm ph}$) is 
smaller in the blackbody spectrum ($dn_\gamma/d\omega\propto\omega$; see 
Equation~\ref{eq:BB}) than in the modified blackbody spectrum
($dn_\gamma/d\omega\propto\omega^0$; see Equation~\ref{eq:modBB}).
\eng{Consequently}, at large radii, the radiative force via resonant scattering 
is weaker in the usual blackbody spectrum than in the modified blackbody 
spectrum resulting in the lower \wada{four-}velocity in the model {\tt BB} than in the
model {\tt fiducial}. However, the photon index difference of 1 does not 
make a significant difference 
\eng{in} the terminal \wada{four-}velocity \eng{of these models.}

The terminal \wada{four-}velocity of the model {\tt B\_1e16} 
\eng{exceeds} that of the model {\tt fiducial}.
In the model {\tt B\_1e16}, the cyclotron energy is higher at a given radius 
\eng{than in the model {\tt fiducial}}.
\eng{Thus,} photons with higher energies are resonantly scattered \eng{at the radius},
and the higher specific luminosity contributes to the radiative force 
(see Equations~\ref{eq:modBB}). As a result, the higher terminal \wada{four-}velocity is 
realized in the model {\tt B\_1e16}. The stronger the magnetic field, the higher
the terminal \wada{four-}velocity.

In the model {\tt B\_1e16}, the \wada{four-}velocity deviates from
$\Gamma\beta\propto r$ in \eng{the range} $5\lesssim r/r_0\lesssim 30$.
This 
\eng{deviation occurs due to the suppression of } the scattering cross section 
\eng{for} X-mode photons 
\eng{with} $\omega\ll \omega_B$ 
\eng{where} $\sigma\sim \sigma_{\rm T}(\omega/\omega_B)^{-2}$ (see Equations~\ref{eq:sXX}
and \ref{eq:sXO}). The condition $\omega< \omega_B$ breaks around 
$r_{\rm res}/r_0\sim 34\,B_{16}^{1/2}(\hbar\omega)_{0,100\,{\rm keV}}^{-1/2}$ 
(see Equation~\ref{eq:rres} and 
\eng{lower} vertical blue line in Figure~\ref{fig:u_comp}), where the \wada{four-}velocity 
\eng{reverts to} the relation $\Gamma\beta\propto r$.\footnote{
If the radiative acceleration is 
\eng{efficient}, the \wada{four-}velocity at a radius, $\Gamma\beta\sim \Gamma_{\rm ph}r/r_{\rm ph}$, 
is 
\eng{entirely} determined only by the photon field. 
\wada{As long as radiation dominates the dynamics, the fluid is accelerated to 
a speed at which the photon field is isotropic in the comoving frame of plasma.}
\eng{Since} the photon field 
\eng{across} radii is mainly determined \eng{by conditions} around the Thomson 
photospheric radius,
the radius-velocity relation under the radiative acceleration is also determined 
\eng{by this region.}
When the radiative force 
\eng{becomes significant} again at $r/r_0\sim 30$, the \wada{four-}velocity again 
returns to the value $\Gamma\sim \Gamma_{\rm ph}(r/r_{\rm ph})$  with a 
\eng{steeper} acceleration than $\Gamma \propto r$.} 
The relation $\Gamma\beta\propto r$ in $r\lesssim 5$ would be sustained by the 
high photon flux in the inner region. The suppressed
cross section at $\omega<\omega_B$ alters the acceleration history 
but does not 
\eng{affect} the terminal \wada{four-}velocity.

The terminal \wada{four-}velocity of the model {\tt n\_-2} 
\eng{surpasses} that of the model {\tt fiducial} 
\eng{for} the same reason in the model {\tt B\_1e16}:
the magnetic field in the model {\tt n\_-2} is stronger than in the model 
{\tt fiducial} at a given radius. The magnetic field at the surface is the 
same as that in the model {\tt fiducial}, and the suppression on the 
cross section shown in the model {\tt B\_16} is not significant.
\eng{Consequently} the \wada{four-}velocity evolution does not significantly deviate 
from the relation of $\Gamma\beta\propto r$. 
\eng{A} weaker the radial dependence of the magnetic field
\eng{results in a} 
higher the terminal \wada{four-}velocity.

\begin{figure}[tb]
  \begin{center}
          \includegraphics[width=0.7\textwidth]{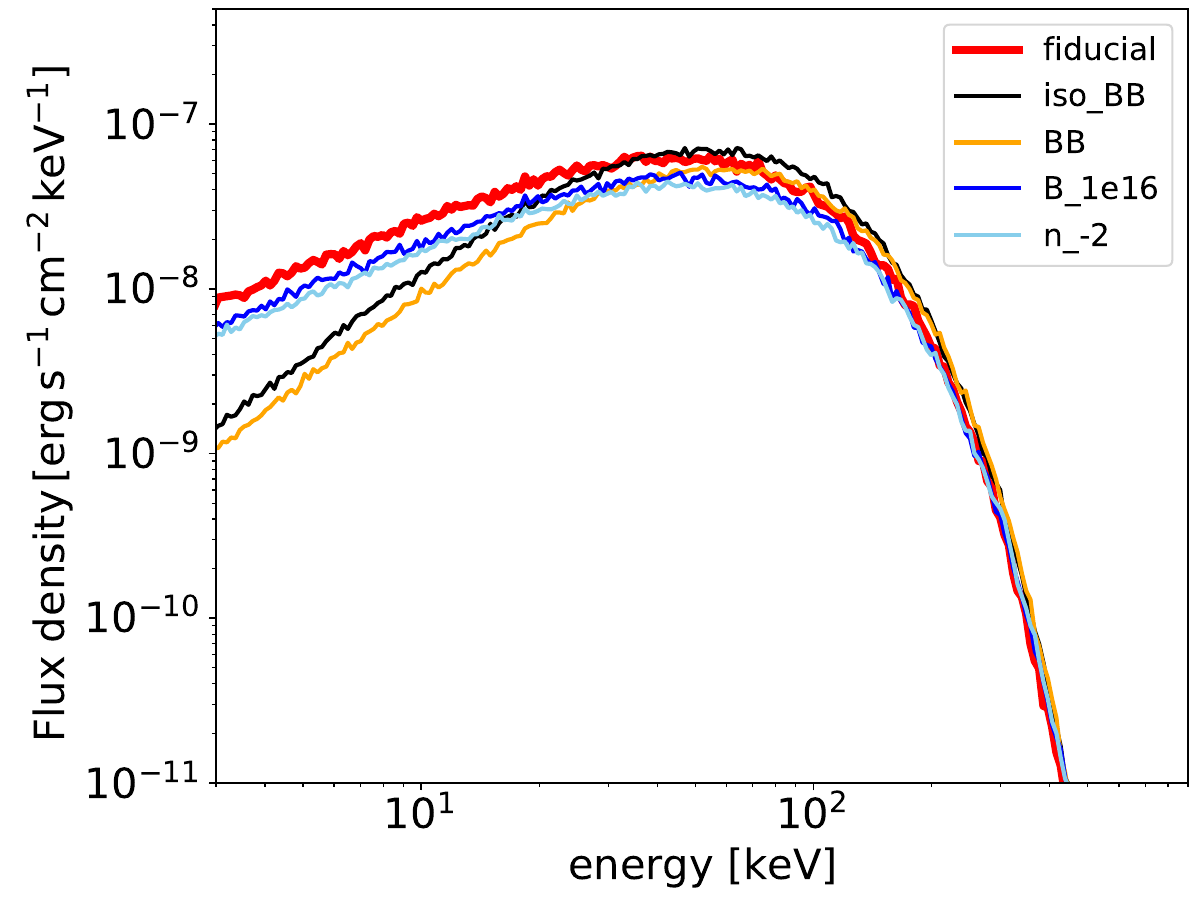}
          \caption{
            X-ray spectrum in 5--900 keV for all the models.
            The axes are the same as Figure~\ref{fig:spec}, and the 
            colors are the same as Figure~\ref{fig:u_comp}.
          }
          \label{fig:spec_comp}
  \end{center}
\end{figure}

Figure~\ref{fig:spec_comp} shows the parameter dependencies of the X-ray
spectrum (see Table~\ref{tab:cases}). In the models {\tt BB} and {\tt iso\_BB}, 
\eng{the blackbody spectrum leads}
the \wada{spectral shapes}
\eng{that are} significantly different from the other models.
Thus, the observed X-ray spectrum in Figure~\ref{fig:spec} (yellow shaded region)
\eng{is unlikely to arise form a blackbody spectrum within this framework.}
The shape of the X-ray spectrum does not strongly depend on 
$B_0$ and $n_B$.

\subsection{Analytic Estimates}\label{sec:ana}
In this subsection, we 
\eng{present} analytic estimates to understand the results of the 
numerical simulations. In Sections~\ref{sec:crT} -- \ref{sec:crrm},
we estimate the coasting radius, $r_{\rm c}$, where the radiative acceleration ends.
We compare the \eng{analytically derived} coasting radii 
\eng{with those inferred} from the simulation results,
\eng{defined by the condition} $d\wada{(\Gamma\beta)}/d(r/r_0)=1$.
The analytical estimates agree with the numerical results within a factor of two.
\eng{By redefining} the coasting radius 
\eng{using the condition} $d\wada{(\Gamma\beta)}/d(r/r_0)=0.3$, 
the analytic estimates agree with the numerical result within \eng{approximately} $15\%$.
Since the accelerated plasma follows $\Gamma\sim \Gamma_{\rm ph}r/r_{\rm ph}$,
in the optically thin regime, the terminal Lorentz factor, $\Gamma_{\rm t}$, is 
approximately equal to $\Gamma_{\rm ph}r_c/r_{\rm ph}\sim 1.5\times (r_{\rm c}/r_0)$.
In Section~\ref{sec:rtau}, we estimate the optical depth for the resonant scattering.

\subsubsection{Thomson Scattering}\label{sec:crT}
\eng{Initially}, we 
\eng{analyze} the case of isotropic scattering with the Thomson cross section.
The radiative force accelerates the outflow if the work \eng{done} by the radiation
during the dynamical time is higher than the rest-mass energy of an electron\cite{MesLag1993,GriWas1998}.
\eng{This} condition
for the radiative acceleration 
is \eng{expressed} as
\begin{eqnarray}
    \frac{\sigma_{\rm T}L_{\rm iso}}{4\pi c\Gamma^2 r^2}\times \frac{r}{\Gamma}>mc^2,
    \label{eq:thomsoncond}
\end{eqnarray}
where $L_{\rm iso}$ is the isotropic photospheric luminosity.
During the radiative acceleration, the Lorentz factor is proportional to 
the radius (see Figures~\ref{fig:u_temp} and \ref{fig:u_comp}) 
\citep{MesLag1993,GriWas1998,LiSar2008}
\begin{eqnarray}
  \Gamma=\Gamma_{\rm ph}\frac{r}{r_{\rm ph}}.
  \label{eq:gammar}
\end{eqnarray}
Substituting Equation~(\ref{eq:gammar}) into Equation~(\ref{eq:thomsoncond}) and 
setting $\Gamma_{\rm ph}/r_{\rm ph}\sim 1$ for simplicity, the coasting radius 
$r_{\rm c}$ is estimated as \citep{NakPir2005}
\begin{eqnarray}
  \frac{r_{\rm c}}{r_0}&\simeq& \left(\frac{L_{\rm iso}\sigma_{\rm T}}{4\pi c^3 m r_0}\right)^{1/4}\sim  12 L_{\rm iso,40}^{1/4}r_{0,6}^{-1/4}.
  \label{eq:rc_bbiso}
\end{eqnarray}
In Figure~\ref{fig:u_comp}, the black dashed line (the model {\tt iso\_BB}) deviates
from \eng{the relation} $\Gamma\beta\propto r$ at $r_{\rm c}/r_0\sim 6.7$
\eng{as} estimated with \eng{the condition} $d\wada{(\Gamma\beta)}/d(r/r_0)=1$.
If we 
\eng{employ} a ``modified'' condition 
$d\wada{(\Gamma\beta)}/d(r/r_0)= 0.3$, \eng{the coasting radius in the simulation is} $r_{\rm c}/r_0\sim 13$.
The terminal \wada{four-}velocity of the fluid, which we define as the 
\wada{four-}velocity at $r/r_0=10^3$, is $\Gamma\beta\simeq16$.
Equation~(\ref{eq:rc_bbiso}) 
\wada{is independent of the magnetic field
and the initial temperature of the fireball,}
$T_{\rm ph}$, \eng{owing to the constancy of the Thomson cross section with respect to the photon frequency.}

\subsubsection{Resonant Scattering with Blackbody Spectrum}\label{sec:bla}
In the case of the blackbody spectrum, the condition for the radiative acceleration 
\eng{via} the resonant scattering is written as (WI23)
\begin{eqnarray}
  \frac{\pi e^2(L_{\omega,\rm iso})_{\omega_B}}{2mc^2r\Gamma^3}>mc^2,
  \label{eq:radacc}
\end{eqnarray}
where $(L_{\omega,\rm iso})_{\omega_B}=\left.dL_{\rm iso}/d\omega\right|_{\omega=\omega_B}$
is the isotropic specific luminosity of the radiation. 
By substituting Equations~(\ref{eq:Br}) and (\ref{eq:gammar}) into Equation~(\ref{eq:radacc})
and using $T\propto r^{-1}$, where $T$ is the temperature at the comoving frame of the fluid\footnote{
We assume that the shape of the photon spectrum depends only on $\omega/T$, \eng{with no additional parametes.}
\eng{Consequently, the adiabatic cooling of freely streaming photons ensures that}
the temperature in the expression of the photon spectrum is proportional to $r^{-1}$ 
\citep[see][for details]{LiSar2008}.}, we estimate the coasting radius.

For the blackbody spectrum, the isotropic specific luminosity is 
\begin{eqnarray}
 (L_{\omega,\rm iso})_{\omega_B}
  &=&L_{\rm iso}\frac{15}{\pi^4}\left(\frac{\hbar\omega_B}{T}\right)^3\frac{\hbar}{T}\frac{1}{\exp(\hbar\omega_B/T)-1}.
  \label{eq:Liso}
\end{eqnarray}
Substituting Equations~(\ref{eq:Br}), (\ref{eq:gammar}), and (\ref{eq:Liso}) into
Equation~(\ref{eq:radacc}) and using $T\propto r^{-1}$ and $\exp(z)-1\simeq z$ for $z\ll 1$,
the coasting radius is (WI23)
\begin{eqnarray}
  \frac{r_{\rm c}}{r_0}&\simeq&
  \left[\frac{15}{2\pi^3}\frac{\hbar e^2}{m^3 c^6 r_0}L_{\rm ph}\frac{\bar{B}_0^2}{\bar{T}_{\rm ph}^3} \left(\frac{r_{\rm ph}}{r_0\Gamma_{\rm ph}}\right)^{3}\right]^{\frac{1}{1-2n_B}},\nonumber\\
    &\sim& 45\times L_{\rm ph,40}^{1/7} B_{0,14.3}^{2/7} \bar{T}_{\rm ph,-1.5}^{-3/7}~~~~(\text{for model {\tt BB}}),\nonumber\\
  \label{eq:rc_BB}
\end{eqnarray}
where $\bar{B_0}=B_0/B_Q$, $\bar{T}_{\rm ph}=T_{\rm ph}r_{\rm ph}/(mc^2 r_0)\sim T_{\rm ph}/(mc^2)$, 
$B_Q=m^2c^3/(\hbar e)=4.4\times 10^{13}\,{\rm G}$, and we set $r_{\rm ph}^{3/7}/(r_0\Gamma_{\rm ph})^{3/7}\sim 1$ 
at the last line for simplicity.
In Figure~\ref{fig:u_comp}, the orange line (the model {\tt BB}) 
\eng{diverge} from $\Gamma\beta\propto r$ at $r_{\rm c}/r_0\sim 38$
($52$ with the modified condition), \eng{which
roughly agrees with the analytical estimation in Equation~(\ref{eq:rc_BB}).}
The terminal \wada{four-}velocity is $\Gamma\beta=94$.
\eng{The coasting radius strongly depends on 
$n_B$ (see Equation~\ref{eq:Br}).}
\eng{For a larger $|n_B|$, $\omega_B$ decreases more rapidly for increasing radii, 
leading to a lower specific luminosity at a given radius (see Equation~\ref{eq:Liso}).  
As a result,} the radiative force decreases more steeply with radius.
\eng{Consequently,} a larger $|n_B|$ 
\eng{results in} a smaller $r_{\rm c}$, equivalently a lower terminal Lorentz factor.

\subsubsection{Resonant Scattering with Modified Blackbody}\label{sec:crrm}
The models {\tt fiducial}, {\tt B\_1e16}, and {\tt n\_-2} 
\eng{employ} the modified blackbody spectrum, for which the isotropic 
specific luminosity is 
\begin{eqnarray}
 (L_{\omega,\rm iso})_{\omega_B}
  &=&L_{\rm iso}\frac{0.073}{T^4}\frac{(\hbar\omega_B)^3}{\exp\left[\frac{(\hbar\omega_B)^2}{T\sqrt{(\hbar\omega_B)^2+(3\pi^2/5)T^2}}\right]-1},\nonumber\\
  \label{eq:Liso_mBB}
\end{eqnarray}
where the \eng{normalization} factor 0.073 
\eng{ensures} $\int d\omega L_{\omega,\rm iso}=L_{\rm iso}$.
By substituting Equations~(\ref{eq:Br}), (\ref{eq:gammar}), and (\ref{eq:Liso_mBB}) into 
Equation~(\ref{eq:radacc}) and using $T\propto r^{-1}$ and $z\ll 1$, we obtain 
\begin{eqnarray}
  \frac{r_{\rm c}}{r_0}
  &=&\left[
\frac{0.073\sqrt{3}\pi^2}{2\sqrt{5}}\frac{\hbar e^2}{m^3 c^6}\frac{L_0}{r_0}\bar{B}_0\bar{T}_{\rm ph}^{-2}
\left(\frac{r_0}{r_{\rm ph }}\right)\left(\frac{r_{\rm ph}}{r_0 \Gamma_{\rm ph }}\right)^4
\right]^{1/(2-n_B)}\nonumber\\
  &\sim&
  \left\{
  \begin{array}{ll}
   90\times L_{\rm ph,40}^{1/5}B_{0,14.3}^{1/5}\bar{T}_{\rm ph,-1.5}^{-2/5}~~&(\text{for model {\tt fiducial}})\\
   280 \times L_{\rm ph,40}^{1/4}B_{0,14.3}^{1/4}\bar{T}_{\rm ph,-1.5}^{-1/2}~~&(\text{for model {\tt n\_-2}})
  \end{array}
  \right.,\nonumber\\
  \label{eq:rc_mBB}
\end{eqnarray}
where we set $[r_{\rm ph}^3/(r_0^3\Gamma_{\rm ph}^{4})]^{1/(2-n_B)}\sim 1$ 
at the last line for simplicity.
In Figure~\ref{fig:u_comp}, the red dashed line (the model {\tt fiducial}) deviates 
from $\Gamma\beta\propto r$ at $r_{\rm c}/r_0\sim 53$ (\eng{or} $83$ with the modified condition)
and the sky blue line (the model {\tt n\_-2}) 
\eng{diverges at} $r_{\rm c}/r_0\sim 140$ (or $260$ with the modified condition).
The terminal velocities are $\Gamma\beta=120$ in the model {\tt fiducial} and $\Gamma\beta=310$ 
in the model {\tt n\_-2}. For the model {\tt B\_1e16}, the analytical estimate
\eng{yields} $r_{\rm c}/r_0\sim 200$
\eng{while} the numerical result shows $r/r_0\sim 120$ (or $190$ with the modified condition).
The terminal \wada{four-}velocity is $\Gamma\beta=250$. These analytic estimates roughly agree with the 
numerical simulations. The soft spectrum in the modified blackbody spectrum 
\eng{enhances the} radiative force at a larger radiii (see Section~\ref{sec:par} for details). 

\subsubsection{Optical Depth for Resonant Scattering}\label{sec:rtau}
The optical depth for the resonant scattering is estimated as follows.
The scattering cross sections for X-mode and O-mode photons close to
the resonance condition, $\mathfrak{u}\sim \wada{\mathcal O}(1)$, are
(see Equations~\ref{eq:sXX}--\ref{eq:sOO})
\begin{eqnarray}
\sigma^X&\simeq& \frac{\sigma_{\rm T}}{2\gamma_e/\pi}\frac{\bar{\gamma}_e/\pi}{(1-\mathfrak{u}^{1/2})^2+\bar{\gamma}_e^2},\label{eq:sXres}\\
\sigma^O&\simeq& \frac{\sigma_{\rm T}\cos^2\theta}{2\gamma_e/\pi}\frac{\bar{\gamma}_e/\pi}{(1-\mathfrak{u}^{1/2})^2+\bar{\gamma}_e^2},\label{eq:sOres}
\end{eqnarray}
where the integration with respect to $d\Omega'$ is performed and 
$\sigma^A \equiv \sum_B\sigma^{AB}$ ($A,B=X,O$).
\wada{It is worth noting here that $\gamma_e$ and $\bar{\gamma}_e$
are the radiative damping factors, which may be less than 1.\footnote{
For $\omega=\omega_B$, the radiative damping factor is 
$\gamma_e\simeq B/(9\times 10^{16}\,{\rm G})$, which is much smaller than 
unity for the parameters in this paper.}}
For a photon 
\eng{of} energy $\hbar \omega$ in the comoving frame of the plasma,
the resonant scattering is significant 
\eng{in the region where}
$|1-\mathfrak{u}^{1/2}|\lesssim \bar{\gamma}_e$ is satisfied (see 
Equations~\ref{eq:sXres} and \ref{eq:sOres}).
The cyclotron frequency $\omega_B$ \eng{depends on the radius and}
can be expanded around $r=r_{\rm res}$ 
(i.e., $\omega_B=\omega$) as $\omega_B\simeq \omega+\Delta r(d\omega_B/dr)|_{\omega_B=\omega}$, 
where $\Delta r=r-r_{\rm res}$. By substituting this expression 
into $|1-\mathfrak{u}^{1/2}|\lesssim \bar{\gamma}_e$, we obtain
\begin{eqnarray}
|\Delta r|\lesssim \Delta r_{\rm res}:= \frac{\bar{\gamma}_e}{|n_B|}r_{\rm res},\label{eq:deltares}
\end{eqnarray}
where we have used Equation~(\ref{eq:Br}) and $r_{\rm res}$ is given 
by Equation~(\ref{eq:rres}) or (\ref{eq:rresc}). The resonant scattering is significant in 
$r_{\rm res}-\Delta r_{\rm res}\lesssim r\lesssim r_{\rm res}+\Delta r_{\rm res}$.
In this range, 
\eng{the cross sections simplify to} 
$\sigma^X\sim \sigma^O \sim \sigma_{\rm T}/(2\gamma_e\bar{\gamma}_e)$.
The optical depth \eng{in this regime is}
\begin{eqnarray}
    \tau_{\rm res}\sim \frac{\sigma_{\rm T}}{2\gamma_e\bar{\gamma}_e}n_\pm(r_{\rm res}) \frac{2\Delta r_{\rm res}}{\Gamma(r_{\rm res})}
    \sim \frac{\sigma_{\rm T}n_\pm(r_{\rm res})r_{\rm res}}{|n_B|\gamma_e\Gamma(r_{\rm res})}.
    \label{eq:taures}
\end{eqnarray}
This is independent of $\bar{\gamma}_e$.
\eng{This invariance arises} because $\bar{\gamma}_e$-dependence of the scattering 
cross section, $\sigma^{X,O}\propto 1/\bar{\gamma}_e$
\eng{is counterbalanced by that of $\Delta r_{\rm res} \eng{\propto}\bar{\gamma}_e$}.
This independence implies that the artificial line broadening by
the parameter $\bar{\gamma}_e$ does not significantly alter 
the number of scattering (see Section~\ref{sec:rad}).

The optical depth $\tau_{\rm res}$ can 
\eng{exceed} unity for X-ray photons.
Using Equations~(\ref{eq:eoc}), (\ref{eq:Br}), (\ref{eq:rres}), 
(\ref{eq:deltares}), and (\ref{eq:taures}),
$\tau_{\rm res}$ during the acceleration ($\Gamma\propto r$) is given by
\begin{eqnarray}
\tau_{\rm res}&\sim& 
\frac{3}{2|n_B|\alpha}\bar{B}_0^{\frac{2}{n_B+1}}\left(\frac{\hbar\omega_o}{mc^2}\right)^{-\frac{n_B+3}{n_B+1}}
\sim 15\,B_{0,14.3}^{-1},
\label{eq:taures_acc}
\end{eqnarray}
where $\alpha$ is the fine-structure constant. We have used 
$\Gamma_{\rm ph}\sim r_{\rm ph}/r_0\sim 1$ for simplicity, and assumed 
$n_B=-3$ in the last line. In the same way, $\tau_{\rm res}$ during the 
coasting ($\Gamma={\rm const.}$) is,
\begin{eqnarray}
\tau_{\rm res}&\sim& 
\frac{3}{2|n_B|\alpha}\bar{B}_0^{\frac{1}{n_B}}\left(\frac{\hbar\omega_o}{mc^2}\right)^{-\frac{n_B+1}{n_B}}\Gamma_{\rm t}^{-\frac{n_B-1}{n_B}} \sim 30\,B_{0,14.3}^{-1/3}(\hbar\omega_{\rm o})_{\rm ,0.1\,keV}^{-2/3}\Gamma_{\rm t ,2}^{-4/3}.
\label{eq:taures_coa}
\end{eqnarray}

\eng{For} the parameters \eng{listed} in Table~\ref{tab:cases}, 
the Thomson photospheric radius is about $r_{\rm ph}\sim r_0$.
Without the resonant scattering, almost all photons escape 
from $r\sim r_0$, \eng{which is} very close to the magnetar surface.
The cross section for X-mode photons with frequency lower than $\omega_B$ is
suppressed by a factor of $\sim (\omega/\omega_B)^2$ (see Equations~\ref{eq:sXX}
and \ref{eq:sXO}). However, this suppression may not \eng{significantly} change 
$r_{\rm ph}$, 
since $r_{\rm ph}$ is just above the magnetar surface.
After the free streaming from $\sim r_{\rm ph}$ to $\sim r_{\rm res}-\Delta r_{\rm res}$, 
the X- and O-mode photons are resonantly scattered \wada{up to} 
$r_{\rm res}+\Delta r_{\rm res}$. 
Therefore, the last-scattering surface for a photon is 
\eng{approximately at} $r_{\rm res}$, 
\eng{which} depends on the frequency of the photon, the magnetic field, and the \wada{four-}velocity 
of the fluid (see Equations~\ref{eq:rres} and \ref{eq:rresc}).

\section{Conclusions and Discussion}\label{sec:con}
In this paper, we numerically study the X-ray radiation, the plasma outflow, and their coupling 
via the cyclotron resonant scattering in outflowing fireballs responsible for magnetar bursts.
For simplicity, we assume a spherically symmetric steady outflow. The plasma is treated as an
anisotropic plasma 
\eng{due to} strong synchrotron cooling (Section~\ref{sec:hyd}). 
Photon propagation is solved in the magnetized electron-positron pair plasma 
(Section~\ref{sec:rad}). The scattering cross section is suppressed 
\wada{by a factor of $\sim (\omega/\omega_B)^2$}
\eng{for} $\omega<\omega_B$ (except in the case that an O-mode photon 
scatters into another O-mode).
Photons of both the two modes satisfying $\omega\sim \omega_B$ are
resonantly scattered. We solve the plasma outflow and the radiation iteratively
\eng{to} find the steady solution 
\eng{for} the plasma and radiation.

We apply our model to the X-ray short burst from a magnetar associated with the 
Galactic FRB, FRB 20200428A (Section~\ref{sec:radacc}). The X-ray spectrum is modified
by Compton 
\wada{scattering} in the relativistic outflow
\wada{and high-latitude emission from a relativistic outflow.}
The resultant spectrum \wada{is}
within 
the \wada{3-sigma} error range of the observed one by the Konus-{\it Wind} (Figure~\ref{fig:spec}). 
The plasma outflow is accelerated by the radiative force, 
\eng{reaching a} four-velocity 
\eng{of} $\sim 100$ \wada{for our fiducial case} (Figure~\ref{fig:u_temp}).
This \wada{four-}velocity is two orders of magnitude higher than the \eng{\wada{four-}velocity of an} outflow without the 
acceleration 
\eng{above $r_{\rm ph}$} and 
one order of magnitude higher than the \eng{\wada{four-}velocity of an} outflow without 
cyclotron resonant scattering.

\eng{We also perform a comprehensive numerical and analytical study of radiative 
acceleration via resonant scattering (Sections~\ref{sec:par} and \ref{sec:ana}).}
We showed that the strong magnetic field enhances the terminal \wada{four-}velocity of the outflow.
\eng{Additionally,} the initial spectrum does not strongly change the terminal \wada{four-}velocity 
within the models of our simulation (Figure~\ref{fig:u_comp}).
These numerical results are roughly consistent with our analytic estimate
(Section~\ref{sec:ana}).

Our numerical simulations show that a relativistic outflow with $\Gamma\sim 100$
is always
\eng{formed} in the outflowing fireball near the magnetic pole.
Such relativistic outflows can 
\eng{serve as} the source for Galactic FRB 20200428A, \eng{which is} associated 
with the magnetar short burst.
\wada{Although the fireball in this study is composed of 
purely electron-positron pairs, in fireballs}
\eng{containing} baryons, relativistic outflows with $\Gamma\sim100$ is 
\wada{also} analytically predicted 
\eng{by} WI23. The kinetic luminosity of 
\eng{such} relativistic outflow can be 1--$10^3$ times higher than the observed FRB 
luminosities (Figure~6 in WI23). Baryons in fireballs can also 
\eng{alter} the polarization of X-ray bursts because the normal modes and the 
cross sections
\eng{differ between} 
pair plasma and 
baryonic plasma.
Circular polarization of the X-ray bursts 
could provide 
\eng{insight into} the amount of baryons and 
\eng{potentially} the trigger mechanism of the bursts \citep{WadShi2024}.
Polarization \eng{may thus serve as a key discriminator between X-ray
emission models associated with FRBs \citep{ZhoLi2024}.}
Motivated by these discussios, the baryon-loaded fireball will 
be investigated in our next paper.

In our simulations, the curvature effect of the magnetic field lines 
is not taken into account. 
\wada{Assuming a dipole magnetic field, the radius $r$ and the polar angle
$\Theta$ satisfy $\sin\Theta\propto r^{1/2}$ along the magnetic field lines.
In the proximity of the magnetic poles, where $\Theta$ is much smaller than unity,
$\Theta$ scales as $r^{1/2}$ along the field lines. 
If the causally connected region of the fireball, 
$r/\Gamma$, exceeds the total size of the fireball, $r\Theta$, 
the curvature effect is insignificant.
To ensure $r\Theta<r/\Gamma$ at $r/r_0=100$ and $\Gamma=100$,
$\Theta<10^{-3}$ is required at $r/r_0=1$.
This value is smaller than the initial size of the fireball 
in this simulation 
(see Table.~\ref{tab:cases}).}
This motivates future studies with 2D simulations.
The outflow near the outer edge of the fireball 
can be decelerated by the radiation \citep{Bel2013}
\eng{potentially producing a harder spectrum} than our results due to scatterings in 
a shear flow \citep{LunPee2013,ItoNag2013,ItoNag2014}. 
\eng{Additionally,} the radial dependence of $T_\parallel$ is also altered when 
the geometry of the magnetic field is different,
\eng{which may further affect} the X-ray spectrum.
We need 2D simulations to include this curvature effect.
\wada{The structured angular profile,
which are
neglected in this paper, may make the spectrum closer to the observed one
much more as studied in \cite{LunPee2013,ItoNag2013,ItoNag2014} for GRB cases.}
This subject is also left as our future studies.


\section*{Acknowledgment}
We thank  K. Ioka, K. Kashiyama, K. Murase, W. Ishizaki, K. Kawaguchi, T. Kawashima, H. Ito, 
and A. Suzuki for fruitful discussions and comments. We also thank the anonymous referee \eng{of PTEP} for 
helpful comments and suggestions. We thank R. Goto, Y. Ichinose, T. Kinugawa, Y. Kusafuka, 
K. Nishiwaki, T. Ohmura, and J. Shimoda for daily discussions. 
This work is supported by 
Grants-in-Aid for Scientific Research, Nos. 22K20366 and 23H04899 (T.W.), Nos. 22K03684, 23H04899, and 24H00025 (K.A.), 
from the Ministry of Education, Culture, Sports, Science and Technology (MEXT) of Japan.
T.W. acknowledges the support of the National Science and Technology Council of Taiwan through grants (113-2112-M-005-009-MY3, 113-2123-M-001-008-, and 111-2112-M-005-018-MY3) and the support of the Ministry of Education of Taiwan (113RD109).
Our study is also supported by the joint research program of the Institute for Cosmic Ray
Research (ICRR), the University of Tokyo. Numerical computation in this work was carried 
out at the Yukawa Institute Computer Facility.


\bibliographystyle{ptephy}
\bibliography{cite}{}
%

%
%
%


\appendix

\section{Derivation of Radiative Force}\label{sec:der}
Here, we derive the radiative force in Equation~(\ref{eq:radforce}).
We assume that electrons are cold, no absorption nor emission occurs, the cross section is given by Equations~(\ref{eq:sXX})--(\ref{eq:sOO}), and the spacetime is flat.
In this part, the photon distribution function $F_\gamma(x^\mu,{\mbox{\boldmath $p$}})$ is denoted as $F(x,p)$ and the unit with $c=1$ is used for
notation simplicity.

The radiative force is written as \citep{Lin1966}
\begin{eqnarray}
   G_{\rm rad}^\nu&=&\int d\Pi\,p^\nu \left(\frac{dF}{d\lambda}\right)_{\rm coll}.
\end{eqnarray}
Using Equation~(\ref{eq:scat}), this is rewritten as 
\begin{eqnarray}
    G_{\rm rad}^\nu&=&\frac{\rho(x)}{m}\int d\Pi\,p^\nu \left[\int d\Pi^\prime\,\kappa(x,p^\prime)\zeta(x;p^\prime\to p)F(x,p^\prime)-\kappa(x,p)F(x,p)\right].  \label{eq:appG}
\end{eqnarray}
This expression is valid in any coordinate system.
For elastic scatterings, the invariant phase function in the fluid rest frame is \citep{Lin1966}
\begin{eqnarray}
  \zeta(x;p^\prime\to p)=\frac{P_{\rm ang}(\cos\theta,\cos\theta^\prime)}{4\pi}\frac{\delta(p^{\prime~(t)}-p^{(t)})}{p^{(t)}},
  \label{eq:appzeta}
\end{eqnarray}
where $\theta$ is the angle between the magnetic field and the wavevector of the scattered photon in the electron rest frame,
$\theta^\prime$ is that of the incident photon,
and $P_{\rm ang}(\cos\theta,\cos\theta^\prime)$ is the angular dependence of the scattering angle, which satisfies $\int d\Omega P_{\rm ang}(\cos\theta,\cos\theta')/(4\pi)=1$.
The superscript $(t)$ represents the 
time component of the four-momentum in the fluid comoving frame.
Substituting Equation~(\ref{eq:appzeta}) into Equation~(\ref{eq:appG}), 
the first term in the square brackets on Equation~(\ref{eq:appG}) is
\begin{eqnarray}
  & &\int d\Pi\,p^\nu \int d\Pi^\prime\,\kappa(x,p^\prime)\zeta(x;p^\prime\to p)F(x,p^\prime)\nonumber\\
  &=&  \int d\Pi\,p^\nu \int d\Pi^\prime\,\kappa(x,p^\prime)F(x,p^\prime)\frac{P_{\rm ang}(\cos\theta,\cos\theta^\prime)}{4\pi}\frac{\delta(p^{\prime~(t)}-p^{(t)})}{p^{(t)}}\nonumber\\
  &=&  \int p^{(t)}dp^{(t)}d\Omega d\Omega'\,\kappa(x,p^\prime)F(x,p^\prime)
  p^\nu \frac{P_{\rm ang}(\cos\theta,\cos\theta^\prime)}{4\pi}.\label{eq:appdp}
\end{eqnarray}
In the comoving frame of the fluid, we can set $p^{(\nu)}=p^{(t)}(1,\sin\theta\cos\phi,\sin\theta\sin\phi,\cos\theta)$ without losing generality because $p^{(t)},\, \theta$, and $\phi$ are the integral variables.
If $P_{\rm ang}(\cos\theta,\cos\theta^\prime)$ is even function of $\cos\theta$ as Equations~(\ref{eq:sXX})--(\ref{eq:sOO}),
\begin{eqnarray}
 \int d\Omega  p^{(t)}\frac{P_{\rm ang}(\cos\theta,\cos\theta^\prime)}{4\pi} &=&p^{(t)}=-(p^\mu u_\mu),  \\
 \int d\Omega  p^{(i)}\frac{P_{\rm ang}(\cos\theta,\cos\theta^\prime)}{4\pi} &=&0~~~~~(i=1,2,3),
\end{eqnarray}
where we have used the normalization condition on $P_{\rm ang}(\cos\theta,\cos\theta')$.
The Equation~(\ref{eq:appdp}) in the comoving frame is rewritten as
\begin{eqnarray}
  \int d\Pi\,p^{(\nu)} \int d\Pi^\prime\,\kappa(x,p^\prime)\zeta(x;p^\prime\to p)F(x,p^\prime)  
  &=&\int p^{(t)}dp^{(t)} d\Omega'\,\kappa(x,p^\prime)F(x,p^\prime)  (-p^\mu u_\mu)\delta^{(\nu)}_{(t)},\nonumber\\
  &=&\int d\Pi\,\kappa(x,p)F(x,p)  (-p^\mu u_\mu)u^{(\nu)},\label{eq:appdpl}
\end{eqnarray}
where we have rewritten the integral variables for angles to be unprimed.
Using Equations~(\ref{eq:appG}) and (\ref{eq:appdpl}), the radiative force in the comoving frame is
\begin{eqnarray}
G_{\rm rad}^{(\nu)}&=&\frac{\rho(x)}{m}\int d\Pi\, \kappa(x,p) F(x,p)\left[-p^{(\nu)}-(p^\mu u_\mu)u^{(\nu)}\right],
\end{eqnarray}
and in any frame, the radiative force is written as 
\begin{eqnarray}
G_{\rm rad}^{\nu}&=&\frac{\rho(x)}{m}\int d\Pi \, \kappa(x,p) F(x,p)\left[-p^{\nu}-(p^\mu u_\mu)u^{\nu}\right]\nonumber \\
  &=&-\frac{\rho(x)}{m}h^\nu_\mu\int d\Pi\, p^\mu \kappa(x,p) F(x,p).
\end{eqnarray}

\end{document}